\documentclass[]{aastex}
\usepackage{emulateapj5}

\begin{document}

\newcommand{\etal}{{\em et al.\/}}
\newcommand{\ie}{{\em ie.\/}}
\newcommand{\cmq}{cm{$^{-3}$}}
\newcommand{\per}{$^{\rm{-1}}$}
\newcommand{\tc}{{$\theta^1$~Orionis~C}}
\newcommand{\ta}{{$\theta^1$~Orionis~A}}
\newcommand{\msol}{M{$_{\odot}$}}
\newcommand{\lsol}{L{$_{\odot}$}}
\newcommand{\kms}{km~s{$^{-1}$}}
\newcommand{\hii}{H{\sc ii}}
\newcommand{\Hii}{H{\sc ii}}
\newcommand{\Ha}{\mbox{H$\alpha$}}
\newcommand{\Hb}{\mbox{H$\beta$}}
\newcommand{\ha}{\mbox{H$\alpha$}}
\newcommand{\La}{\mbox{L$\alpha$}}
\newcommand{\oiii}{O{\sc iii}}
\newcommand{\oii}{O{\sc ii}}
\newcommand{\oi} {O{\sc i}}
\newcommand{\nii}{N{\sc ii}}
\newcommand{\cii}{C{\sc ii}}
\newcommand{\ciii}{C{\sc iii}}
\newcommand{\civ}{C{\sc iv}}
\newcommand{\sii}{S{\sc ii}}
\newcommand{\feii}{Fe{\sc ii}}
\newcommand{\Feii}{Fe{\sc ii}}
\newcommand{\si}{S{\sc i}}
\newcommand{\mgii}{Mg{\sc ii}}


\title{ Irradiated Herbig-Haro Jets  \\
        in the Orion Nebula and Near NGC~1333} 

\author{John Bally \altaffilmark{1,3} 
    and Bo Reipurth \altaffilmark{2,3}}

\affil{Center for Astrophysics and Space Astronomy and \\
       Department of Astrophysical and Planetary Sciences, \\
       University of Colorado, Boulder, CO 80309}
\authoraddr{ Campus Box 389, Boulder, CO 80309-0389}

\altaffiltext{1}{bally@casa.colorado.edu}
\altaffiltext{2}{reipurth@casa.colorado.edu}

\altaffiltext{3}{Visiting astronomers at Kitt Peak National Observatory
which is operated by the Association of Universities for Research 
in Astronomy (AURA) Inc. under contract with the National Science Foundation.}

\begin{abstract}
We report the discovery of a dozen Herbig-Haro jets 
illuminated by the Lyman continuum ($\lambda ~<$ 912\AA )
and/or softer far-ultraviolet (912\AA\ $<~ \lambda ~<$ 2000\AA )
radiation fields of nearby high mass stars.  Five irradiated outflows 
lie in the outer parts of the Orion Nebula (HH~502 through HH~506) and seven 
lie near the reflection nebula NGC~1333 in the Perseus molecular 
cloud (HH~333 through HH~336, and HH~497 through HH~499).  
These stellar outflows are powered by optically visible low 
mass young stars which suffer relatively low extinction and
seem not to be embedded within opaque cloud cores.   We propose that
the UV radiation field has eroded residual material left over from 
their formation on a time-scale short compared to the ages of these 
star forming regions.  Many of the irradiated jets exhibit unusual C-shaped 
symmetry.  In the outskirts of the Orion Nebula, most irradiated 
jets appear to bend away from the core of the nebula.  On the other hand, 
in NGC~1333, the C-shaped jets tend to bend back towards the cluster 
center.  Jet bending in the Orion Nebula may be dominated by either 
the outflow of material from the nebular core or by the rocket effect 
pushing on the irradiated portion of a mostly neutral jet beam. 
But in NGC~1333,  jet bending may indicate that the source stars 
have been ejected from the cluster core.  Many irradiated jets are 
asymmetric with one beam much brighter than the other.  When fully 
photo-ionized, 
irradiated  jets may provide unique insights into the physical conditions 
within outflows powered by young stars, permitting the determination 
of the density and location of stellar ejecta even in the absence of 
shocks.  We present a model for the photo-ionization of these outflows by 
external radiation fields and discuss possible mechanisms for producing 
the observed asymmetries.  In particular, we demonstrate that the UV 
radiation field may alter the amount of cloud material entrained by the jet.  
Radiation induced variations in mass loading and beam heating can produce 
differences in the beam velocities and spreading rates, which in turn
determine the surface brightness of the radiating plasma.   
In a bipolar irradiated jet in which both beams have the same mass loss 
rate and opening angle, the {\it slower} beam will appear brighter at 
a given distance from the source.  On the other hand, if both beams 
spread orthogonal to the jet propagation direction with the same speed
(e.g. both beams have the same internal sound speed or shocks with similar 
physical conditions), the {\it faster} beam will appear brighter at the 
same distance from the source.  Thus, depending on the parameters, 
either the faster or slower beam of a jet can be brighter.   
Finally, we report the discovery of some large scale bow shocks which 
face the core of the Orion Nebula and which surround visible young stars.  
These wind-wind collision fronts provide further evidence for a large scale 
mass flow originating near the nebular core.

\end {abstract}


\section{ Introduction}

Young stars, born from dense molecular cloud cores, frequently 
drive highly supersonic jets and winds which are traced by 
Herbig-Haro (HH) objects at visual wavelengths, near infrared emission 
line regions bright in molecular hydrogen and [\feii ], and the ubiquitous 
bipolar molecular outflows observable at millimeter wavelengths.  
HH objects are collisionally excited nebulae produced by outflows 
ejected by young stellar objects (YSOs) and they either trace 
shocks where outflows ram the ambient cloud (terminal working 
surfaces) or shocks produced by colliding fluid elements ejected 
from the source at different velocities and times (internal working 
surfaces).  Usually, they can be identified by 
their forbidden [\sii ] emission and large [\sii ]/\Ha\ line ratio.   
Since Herbig-Haro objects are thought to be produced mainly during 
the first few hundred thousand years of the life of a YSO,  
most HH sources are highly obscured by the cloud cores and
circumstellar environments from which they formed.

The optical emission lines which trace normal HH objects
are excited by non-equilibrium processes in and behind shock 
waves.  Therefore, the intensities of the emitted  lines depend on the 
complex details of the flow such as the shock geometry, the structure and 
strength of magnetic fields, the ionization fraction, and the effects 
of the radiation generated by the shocks themselves on the upstream and 
downstream material.  Therefore,  it is very difficult to reliably 
estimate the basic physical parameters such as density, temperature,
and ionization fraction and how these parameters vary across the emission 
region.  Furthermore,  it is even harder to infer 
the overall parameters that characterize an outflow such as the stellar 
mass loss rate, outflow momentum, and flow kinetic energy flux, since 
moving material which remains quiescent (unshocked) is not 
traced by these emission lines at all.  Therefore, most of the mass 
(and therefore the momentum and kinetic energy) of an outflow may 
remain invisible.

Reipurth \etal\ (1998) recently identified four Herbig-Haro jets 
located near the luminous high mass multiple star system $\sigma$ 
Orionis,  all four of which originate from {\it visible} T-Tauri stars 
{\it not} embedded in opaque molecular cloud cores.  The intense
ultraviolet radiation field of the massive stars has ionized and 
removed the bulk of the interstellar matter from the YSO environment.    
Interestingly, all of these jets are asymmetric, with the beam facing 
away from $\sigma$ Orionis being much brighter than the counter beam.

The discovery of externally irradiated HH objects such as those found by
Reipurth \etal\ (1998) and Cernicharo \etal\ (1998)
creates new opportunities for diagnosing outflows.  
In externally irradiated outflow systems, the intensity of the emission 
produced by a variety of spectral lines may be primarily determined by 
the external illumination rather than the complex physics of shocks.  
For example, if the external radiation field ionizes either the surface 
or the entire body of a jet or wind, the observed intensity of the 
\Ha\  line can be directly related to the emission measure, and 
the red [\sii ] lines can be used to measure the electron density.  
Measurements of the radial velocity field and flow morphology can be 
combined with these parameters to infer many flow properties such as the
mass loss rate, momentum carried in each beam, and the energy injected
into the cloud by the flow.   Instead 
of relying on highly uncertain shock models, we can apply the well 
understood theory of photoionized nebulae to estimate outflow parameters.
Furthermore, external radiation fields are likely to illuminate {\it all} 
unshadowed outflow components equally so that the flow structure can be 
determined reliably.

In this paper we report the discovery of new irradiated jets in two
star forming regions;  the Orion Nebula  (M42, NGC 1976; d = 470 pc) and  
the NGC~1333 region in the Perseus molecular cloud (d = 220 pc).
We first discuss the observations and present a description of
each new flow.  Then we present a simple photo-ionization model
for irradiated jets, discuss a possible mechanism for the
production of jet asymmetries, and finally explore the implications
of C-shaped symmetry observed in jets contained both within the 
Orion Nebula and near the core of NGC~1333.

\section{ Observations}

We obtained narrow band images of the Orion Nebula and
NGC~1333 on the nights of 28$-$30
October 1997 at the prime focus of the Mayall 4 meter
telescope using the MOSAIC 8192 $\times$ 8192
pixel CCD camera equipped with the engineering grade CCD
detectors.  The field of view of this setup is
approximately 36\arcmin\ $\times$ 36\arcmin\ with a scale of
0.26\arcsec\ per pixel (for details, see the MOSAIC homepage at {\it
http://www.noao.edu/kpno/mosaic/mosaic.html}).   To minimize
the effects of bad columns and pixels, and the 40 pixel-wide
gaps between the individual CCD detectors, we obtained for each 
field a set of five 600s exposures using a dither pattern with a 
spacing of 100 pixels.  Images were obtained through an \ha\ 
narrow band filter centered on 6563 \AA\ with a band pass of 75 \AA, 
and through a [\sii ] filter centered on 
6730 \AA\ with a band pass of 81 \AA.  The data were overscanned, 
trimmed, dark subtracted, and flat fielded using the
IRAF package {\sl MSCRED}.  This package compensates for the 
geometrical distortions introduced by the 4 meter optics, 
including the effects of the field flattener.
Dark subtraction was based on a median-combined set of eleven 
600s dark exposures taken at various times throughout the 
observing run.  A comparison of darks taken at different times 
during the run did not show any significant
variations.  Flat fielding was based on sky flats.
The \ha\ flat was formed by median combining nine twilight
sky flats, while the [\sii ] flat was formed by median combining
thirteen sky flats. After reduction, individual images were 
combined into a single FITS image in each filter with IRAF.
An internal reflection resulting from a defective anti-reflection 
coating on the prime-focus field flattener/corrector produced a 
large doughnut shaped artifact in the images that persists at the 
few percent level after flat fielding and median combining.  
Since the amount of reflection depends critically on the location 
of bright objects in the imaged field, its intensity varies from 
image to image.   This artifact is less noticeable in regions with 
strong emission.

Visual wavelength spectra were obtained with the RC spectrograph at
30~\kms\ resolution at the Mayall 4~m reflector at Kitt Peak, on the
nights of December 26 through 30, 1998.  Spectra covering the wavelength 
range from [\oi] $\lambda$6300 to [\sii] $\lambda$6731 were obtained with
a  2\arcsec\ wide and 5\arcmin\ long slit which was oriented in each flow
at the position angle listed in Table~1 using grating KPC 24 in second
order and a long-pass glass order sorting filter (OG 570).   
The image scale of this setup was 0.69\arcsec\ per pixel along
the slit and 0.54\AA\ per pixel (24 \kms ) along the dispersion direction.
The 2\arcsec\ slit provides a spectral resolution of about 0.7\AA\ or
32 \kms .  For further details on the 
instrument and data reduction, see Devine \etal\ (1997).
All radial velocities quoted in this paper are with respect to the
velocities of the diffuse nebular lines produced by M42 and NGC~1333 
which are assumed to trace the stationary ambient medium.

\section{ Results}

\subsection{Irradiated Jets in the Outskirts of the Orion Nebula}


The Orion Nebula (Figure~1) is the nearest region of active high 
mass star formation.  Located at a distance of about 470 pc, the
nebula contains the Trapezium Cluster of O and early B stars
and over 700 low mass stars within a region less than a few parsecs 
in diameter (e.g. McCaughrean \& Stauffer 1994).  The most luminous member, 
\tc\ (O6pe) is responsible for ionizing the \hii\ region. 
The ten parsec long dense filament of molecular gas at the northern 
end of the Orion~A cloud which spawned the Orion Nebula is 
associated with at least 2,000 YSOs, the majority of 
which are less than 1 million years old (Hillenbrand 1997; 
Hillenbrand \& Hartmann 1998).   Bally, O'Dell, \& McCaughrean (2000a)
report the discovery of 24 compact irradiated microjets emerging from 
the `proplyds' (an acronym for PROto PLanetarY Disk) in the nebular core.


Figures 2--6 show [\sii ] and \Ha\ images of the irradiated jets embedded
in the outskirts of the Orion Nebula.   
Four irradiated jets lie in the southern and western parts of the Nebula.
The appearance of these jets is likely to be dominated by
the Lyman continuum (LyC or EUV) radiation field of the Trapezium stars.
One jet lies towards the northeast well beyond the projected boundary of
the nebula where it is likely to be only exposed to softer far ultraviolet
(FUV) radiation.

{\bf HH 502:}  
This outflow (Figure~2) consists of a knotty 60\arcsec\ long jet,
extending south towards at least three bow shocks of
progressively larger size and lower surface brightness that
lie up to 150\arcsec\ from the source.  Although the counterjet 
is more than three times fainter, there are two relatively
high surface brightness bow shocks on the north side of this outflow.
Overall, the jet and the chain of bow shocks bend away from the
Orion Nebula core.

The outflow originates from a 17-th magnitude \Ha\ emission 
line star located 2.8\arcmin\ east and 6.2\arcmin\ south
of the Trapezium (\tc ).   The stellar \Ha\ emission has 
an equivalent width 
(EW) of about 60\AA\ and can be traced to about $\pm$ 700 \kms\ on 
either side of the nebular \Ha\ emission, indicating that the source
of the HH~502 jet is a classical T Tauri star.  Strong forbidden 
[\oi ] emission in the $\lambda$ 6300 and $\lambda$ 6363\AA\ lines 
is seen from the star with EW([\oi ]) $\sim$ 10\AA\ in each line.   
This emission is predominantly redshifted by about 30 \kms\ 
from the nebular [\oi ] emission.   However, faint [\oi ] line wings 
can be traced up to $\pm$ 75 \kms\ from the nebular line
on top of the stellar continuum.   In addition to this spatially 
unresolved component, the [\oi ] emission also exhibits a spatially
extended redshifted ($\sim$ 50 \kms ) feature which in the spectra 
can be faintly traced for about 20\arcsec\ south of the source 
along the direction of the jet visible in the \Ha\ image (Figure 1).

The main jet from this star extends to the south at 
PA = 208$^o$ and is  much brighter in \Ha\  than in [\sii ]
(Table 2A, 3).  Although the jet fades considerably about 5\arcsec\ south 
of the source,  it can be traced for nearly an arc minute to the  
south as a series of about seven knots and faint interconnecting
emission.   Remarkably, the jet remains unresolved in the
transverse direction (width $<$ 1\arcsec ) for up to 
40\arcsec\ from the source!   The jet bends to the south by about
5\arcdeg\ roughly 25\arcsec\ south of the source.   The position
angles of various features in the HH~502 outflow with respect to
the central star, and the deflection from the nominal launch
direction of the jet are listed in Table 2A. 

At least three \Ha\ bow shocks are seen at intervals 
of about 45\arcsec\ along the jet axis south of the source.  
Knots of [\sii ] emission are found at the tips of the bows.
Two \Ha\ dominated bows are
seen towards the opposite side of the source in the counterflow direction
at PA = 40$^o$.   The three bow shocks south of the source and the two lying
north of the source do not lie in a straight line, but on a curve that
indicates C-shaped symmetry about the source.  

Three RC Spectrograph long-slit orientations and placements were used
to observe HH~502.  The first slit (PA = 22\arcdeg\ or 202\arcdeg ) 
is centered on 
the source star and passes near the core of the first compact bow 
shock (S6, located 50\arcsec\ from the source) and the wings of 
the next bow shock (S7, located about 10\arcsec\ downstream).  
Except within 5\arcsec\ of the source, this slit orientation misses 
most of the curved jet which emerges towards the southwest.  
The second slit, (PA = 40$^o$) is also centered on the source star 
and was chosen to pass through the apices of the two bow shocks lying to 
the northeast of the source (N2 and N3).  This slit included the light
of the star superimposed on the northern lobe of HH~502 (star A) and
star B in Figure 1.   Finally, the third 
slit (PA = 7$^o$) is centered on a faint star (star C) in the southwest 
lobe of the HH~502 flow and passes through the brightest knots at the 
apices of the bow shocks S6 and S7 as well as the eastern side of
S8 located 30\arcsec\ downstream from S7.  

The results of the spectroscopic observations of HH~502 show that:
(1) The brighter southern jet is redshifted by +50 \kms \ with respect
to the Orion Nebula with wings extending to about +70 \kms . 
The jet kinematics is most apparent in the 
[\nii ] lines since the \Ha\ line is overwhelmed by the strong
background produced by the Orion Nebula.  However, unlike the \Ha\ emission,
the [\nii ] emission is brighter towards the northern counter jet than 
towards the main jet that extends south of the source.  Faint [\sii ] 
emission is also visible within several arc seconds of the source on both
sides.    
(2) The flow from the young star exhibits kinematic asymmetry with
the fainter northern lobe being blueshifted by about $-$100 \kms\ with
wings extending to about $-$130 \kms .  The jet and counter jet
are more easily seen in [\nii ] than in \Ha\ due to the better contrast 
against the nebular background.  As the [\nii ] lines, the [\sii ] 
emission also shows this kinematic asymmetry.
(3) The blueshifted northern bow shocks N2 and N3 have Doppler shifts of
$-$130 \kms\ and $-$70 \kms\ respectively.
(4) Doppler shifts of only two of the southern bow shocks were measured.
Both S6 and S7 have radial velocities of about +25 \kms\ with
respect to the Orion Nebula lines.
(5) All four stars which fell on the various slit orientations
show evidence of youth.   Star A has strong \Ha\ emission and
lithium $\lambda$6707 in absorption.  Star B, though lacking stellar
\Ha\ emission, exhibits a very deep lithium absorption line, and star
C exhibits very strong \Ha\ emission and weak lithium absorption.
The source of HH~502 does not exhibit a measurable lithium absorption
feature.  Thus, all of these stars are likely members of the extended
Trapezium cluster. 
(6) Finally, slit 2 reveals a $-$75 \kms\ feature, f, visible in \Ha\ and [\nii ] 
about 30\arcsec\ southwest of the driving source of HH~502 which is not 
associated with the HH~502 outflow system.  The \Ha\ image shows a
faint bow shaped emission region here.

{\bf  HH 503:} 
The second most prominent irradiated jet in the outskirts of the
Orion Nebula is HH~503 (Figure~3), located 9.8\arcmin\ west and 2.2\arcmin\ 
south of HH~502  (6.9\arcmin\ W, 8.4\arcmin\ S of \tc ).   
Unlike most other irradiated jets in \Hii\ regions, 
this flow has a relatively large [\sii ] to \Ha\ brightness ratio
and is therefore seen with greater contrast against the background in 
the [\sii ] image.  

The optical bipolar outflow emerges from an \Ha\ emission line star
with strong lithium absorption.  This star also exhibits forbidden
[\oi ] emission.  Although a bipolar jet is visible on both sides of 
the star in both \Ha\ and [\sii ], it is highly asymmetric 
in brightness, with the portion of the jet extending to northwest 
(PA = 285\arcdeg ) from the star being at least four times
brighter than the counterjet.  The jet can be traced to a compact 
[\sii ] dominated bow shock, HH~503~W1, about 65\arcsec\ from the 
source star.  The wings of this bow shock are highly asymmetric, with  
the northern side that faces the core of the Orion Nebula being
much brighter and larger than the other side.   The north rim
of this bow can be traced for 50\arcsec\ back nearly to the source 
star.   The southern rim of this shock disappears within a few
arc seconds of the tip of the bow.   The extended north wing lies 
about 5\arcsec\ north and its curvature closely follows the bent body 
of the jet.  A second and brighter bow shock (HH~503~W2) lies about 
18\arcsec\ farther west.  Unlike HH~503~W1, 
this object does not exhibit extended wings and is only about 4\arcsec\
in size.   The counter jet can be faintly traced towards the 
southeast where it terminates in a compact bow shock (HH~503 E1) about 
45\arcsec\ from the source.  Knot E1 is the only component in this
flow which has a large \Ha\ to [\sii ] brightness ratio (13.0) that 
is consistent with full photo-ionization.   All of the
other features in this flow have ratios ranging from 2 to 4
(Table 3).  About 70\arcsec\ further along the curving outflow 
axis lies the brightest HH objects in this flow, the double knot 
HH~503~E2.  This structure looks like a fragmenting bow 
shock that is bright in both [\sii ] and  \Ha .  Overall, HH~503 is 
a highly collimated jet system with multiple bow shocks whose outflow 
axis appears to be bent by about 40$^o$. 
Thus, the jet and the associated chain of bow shocks exhibit bending 
away from the core of the Orion Nebula (C-shaped symmetry).

Spectra were obtained with three different RC Spectrograph long slit
positions and orientations.  The slit was oriented at PA = 113$^o$
for the first two spectra.  The first spectrum was centered on the
driving source and the slit includes some of the brightest
portions of the jet extending to the west.  The second
spectrum was obtained with the same slit orientation but displaced 
9.8\arcsec\ south so that HH~503~W1 and E1 both fell on the slit.
The third spectrum,  obtained at PA = 120\arcdeg, includes the brighter
bow shocks HH~503 W2 and E2.  

This bent outflow appears to be redshifted towards the northwest and 
blueshifted towards the southeast based on the kinematics of knots
W2 and E1.   However,  neither the jet nor knot W1 were detected in the
spectra, presumably because of the faintness of these features.  
Knot W2 has a redshifted velocity of about +25 \kms\ while knot E1 
is blueshifted at $-$25 \kms\ with faint \Ha\ line wings extending
to about $-$100 \kms\ with respect to the Orion Nebula.  This wing 
emission suggests that the shocks in E1 have speeds of at least
50 \kms\ (Hartigan \etal\ 1994).  Although it lies on the nominally 
blueshifted side of the HH~503 flow, the large and bright double bow 
shock E2 is redshifted, 
with a mean Doppler shift of about +25 \kms\  with faint emission 
extending up to +75 \kms.  
Therefore, it is possible that E2 is not part of the HH~503 system.  
However, if it is, then the flow must be bent not only in the plane 
of the sky, but also along the line of sight.  Furthermore, the four 
bright compact bow shocks that delineate the bent body of this bipolar 
flow cannot be moving ballistically from the source. Instead, the 
emitting material must itself have undergone a large deflection from 
the initial outflow launch direction both in the plane of the sky 
and along the line of sight.  The spectra also demonstrate that the 
three HH objects in this flow (HH~503 E2, W1, and W2) are most prominent 
in the 6300\AA\ [\oi ] and in the [\sii ] lines, and barely detected 
in \Ha , indicating that these HH objects have large neutral components. 

{\bf  HH 504:} 
This object (Figure~4), located 3.2\arcmin\ W and 6\arcmin\ S of \tc ,
consists of a 10\arcsec\ long [\sii ] filament extending 
from near a bright star towards position angle $PA$ = 345$^o$.  
A chain of compact [\sii ] dominated knots 
and a compact partial bow shock lie about 30\arcsec\ to the north. 
A peculiar feature of this jet is that the jet axis seems to miss the 
bright star by about 0.5\arcsec , perhaps because it originates from a
fainter companion.   The brightest [\sii ]  emission 
is displaced from the star towards the core of the Orion Nebula by about 1\arcsec .
This jet appears to be unipolar.
Several faint filaments of [\sii ] emission lie about 30\arcsec\ northeast
of the source.  Is is not clear if these features are associated with
the outflow system or are unrelated objects.   

{\bf  HH 505:} 
Located 8.9\arcmin\ west and 0.7\arcmin\ north of \tc , this one-sided 
jet (Figure~5), most visible in the [\sii ] image extends from a 
star at PA = 345\arcdeg\ and terminates in a bright knot of [\sii ] emission
about 15\arcsec\ away.  A crescent shaped \Ha\ bright rim wraps 
around the star and its unipolar jet on the side of the system 
facing the Orion Nebula core.  The intensity
distribution indicates that this feature may trace the limb brightened 
skin of a bubble that wraps around the star.  The curvature of this arc of
\Ha\ emission is most pronounced near the star, which is also where the 
arc reaches its peak surface brightness.    
The \Ha\ arc is similar to other predominantly \Ha\ 
bright crescents which appear to wrap around dozens of young stars in the 
outskirts of the Orion Nebula (the LL Ori objects; see below).   

{\bf  HH 506:} 
A 4\arcsec\ long jet, located 7.7\arcmin\ east and 
13.0\arcmin\ north of \tc\ (Figure~6), emerges from a bright star at 
PA = 30$^o$ towards the north.  The star is centered on a clumpy 
60\arcsec\ long curved filament of [\sii ] emission that bends towards 
the east on both sides of the star.
The southern portion of this chain contains 8 distinct compact 
($<$ 1\arcsec\ diameter) knots of [\sii ] emission.  The northern part
consists of a faint filament of emission.  HH~506 is located northeast of
the Orion Nebula about 10\arcmin\ east of the OMC2/3 molecular ridge and
is therefore outside the projected edge of the photo-ionized portion of the
Orion Nebula.  

This portion of the Orion region contains many extended filaments of
[\sii ] and \Ha\ emission, and numerous HH objects.  For example,
a bright 10\arcsec\ diameter HH complex lies about 50\arcsec\ to the 
northwest of the HH~506 source.   A fainter HH object lies 70\arcsec\ away
along the same direction.  These objects do not appear to be connected
to the HH~506  jet. 


\subsection{ Wind-Wind Collision Fronts} 
 
All jets in the outer portions of the Orion nebula which exhibit 
C-shaped symmetry bend away from the nebular core, indicating the 
presence of a wind or a force which can bend the jets and their cocoons.  
Evidence for a pervasive outflow of material from the inner nebula is 
provided by stars that  are surrounded by bow shaped emission line features 
pointing towards the Trapezium cluster.  The bows are symmetric about 
a line connecting the associated star to the Trapezium stars and 
range in size from about 20\arcsec\ to several arc minutes.  The 
prototype feature is associated with the star LL Ori located 
1.9\arcmin\ west and 2.7\arcmin\ south of \tc\ (Gull \& Sofia 1979; 
Bally \etal\ 2000a).  The minimum projected separation 
between the associated star and the closest point to the arc (usually 
on a line connecting the star to the Trapezium) ranges from about 
3\arcsec\ to about 20\arcsec .  These [\sii ] and \Ha\ bows appear to 
trace shock fronts at the interface between a wide angle wind emerging from 
the associated young star and a large scale outflow from the core of 
the Orion Nebula.  Bally \etal\ (2000a) present Hubble Space Telescope 
observations of ten wind-wind collision fronts within the inner 
4\arcmin\ core region of the Orion Nebula.  Bally \etal\ provide a 
detailed discussion of the mechanisms responsible for these features.  
The wide-field MOSAIC images reveal additional similar objects in the 
outer portions of the nebula.   The LL Ori type wind-wind collision 
fronts found on these images are listed in Table~5 and shown in 
Figures 14 and 15.

\subsection{Irradiated Jets in NGC~1333}

NGC~1333 (Figure~7) is the most active current site of star formation
within the Perseus molecular cloud complex located
at a distance of about 220 to 350 pc (Herbig \& Jones 1983; C\u{e}rnis 1990).
We adopt 220 pc for the distance to NGC~1333. 
Though the designation NGC~1333 strictly applies to a reflection nebula 
located at the northern end of a dense molecular cloud core, it is
generally applied to the whole active site of star formation that has spawned 
at least 150 stars within the last 2 million years 
(e.g. Lada, Alves, \& Lada 1996), and we use it to designate this entire 
star forming region.  There is a remarkable concentration of highly 
collimated stellar jets in the immediate vicinity 
(within $<$ 5\arcmin\ radius) of the NGC~1333 reflection nebula. All of these
jets appear to be powered by visible stars with very small amounts of 
visual extinction.  The locations of objects discussed below are given
with respect to LkH$\alpha$~270 in the middle of the NGC~1333 reflection 
nebula.  This star is located 0.45\arcmin\ west and 2.2\arcmin\ south
of the B9 star BD+30\arcdeg 549, the brightest star in NGC~1333.
HH~334 through 336 were discovered by Bally, Devine, \& Reipurth (1996).
The objects HH~497 through 499 are new.

{\bf  HH 333:} 
A highly collimated jet emerges from an 18-th magnitude star 
1.3\arcmin\ west and 3.4\arcmin\ north of LkH$\alpha$~270 at 
PA = 254\arcdeg\ (Figure~8). The relatively bright [\sii ] doublet towards 
the star implies an electron density at the base of the jet of 
$n_e = 1.3 \times 10^3$ cm$^{-3}$.  
A faint \Ha\ reflection nebulosity can be traced for up 
to 30\arcsec\ from the source, especially on its southwestern side.
The jet beam can be traced continuously on both sides of the central
star for a total length of at least 3.5\arcmin\ end-to-end.  
The jet remains remarkably straight and highly collimated
within 70\arcsec\ of the source (between knots E3 and W2).  
This section of HH~333 contains three nearly equally spaced knots 
on the east side (E1 -- E3), and two knots on the west side 
(W1, W2) of the source.   The knot pairs E1, W1 and E3, W2 are 
equidistant from the source;  E2 does not appear to have an associated 
feature on the western side of the jet.   The electron densities in knots
E3 and W2 are 100 cm$^{-3}$ and 160 cm$^{-3}$, respectively. 

Although the jet beam is unresolved, the knots E3 and W2 are about 2\arcsec\ 
in diameter, implying that the jet remains confined to an opening angle less than
1.6\arcdeg .   Knots E1 through E3 are redshifted by about +50 \kms , 
while W1 and W2 are blueshifted by about $-$50 \kms .  While the redshifted portion 
of the flow ends at knot E3, there are several additional diffuse bow shocks 
further along the jet axis towards the west.  The beam appears to bend
north at knots W3 and W4, and the fragmented bow shocks W5 and W6
continue in a further bending of the beam towards the north.

{\bf  HH 334:} 
This bipolar jet appears to originate from a bright \ha\ emission
line star with EW(\Ha ) = 140\AA\ located 
3.9\arcmin\ west and 1.0\arcmin\ south of LkH$\alpha$~270 
and north of the bright HH object HH 12 (Figure~9).  The jet is knotty, and 
towards the west extends from the star at position angle PA = 290\arcdeg, 
terminating in a bright \Ha\ emission line knot, HH 334 W1.  
Several arc seconds to the south of this knot and of the main body of the jet, 
there is a parallel strand of [\sii ] emission.  The eastern portion of
the HH~334 jet emerges from the star at PA = 120\arcdeg, which deviates by 
about 10\arcdeg\ from the orientation of the western jet. The eastern 
jet is \ha\ bright until it reaches the first knot E1, which is
placed symmetrically with respect to W1 about the suspected source star.
Beyond E1, the jet becomes a [\sii ] bright filament that appears to terminate
in a complex of knots about 1\arcmin\ to the southeast of the source star.

The spectra show that both W1 and E1 are low velocity features with Doppler
shifts of only +25 \kms\ for knot W1 and $-$25 \kms\ for knot E1.  However, 
faint \Ha\ emission can be traced to about +50 \kms\ in knot W1 and the 
line appears to be at least 50 \kms\ wide at low levels.  
Knot W1 is also quite bright in [\nii ] and faintly visible in [\oi ]. 

{\bf  HH 335:} 
HH~335 (Figure~10) is a monopolar bent jet which is nearly 1\arcmin\ long
and which is deeply embedded in the bright reflection nebula surrounding
the star LkH$\alpha$~270 in NGC~1333.  Therefore, it is best seen in 
the continuum subtracted images.  The jet contains several knots along 
its length and gets progressively fainter with increasing distance from 
the apparent source.  At its base, the jet is aligned with the axis 
of symmetry of the reflection nebula and is propagating nearly due north.   
However, the jet bends westward in a graceful arc by about 25\arcdeg\ 
at its northern end. 

{\bf  HH 336:} 
This is a remarkable bipolar jet located symmetrically about an 
\Ha\ emission line star to the east of the main NGC 1333 cloud core
4.7\arcmin\ east and 5.1\arcmin\ south of LkH$\alpha$~270
(Figure~11).
The jet consists of several distinct segments separated by 
emission free gaps that emerges from the star at PA = 135\arcdeg\ or
315\arcdeg.  The northwestern jet segments appear to be more diffuse and
extended orthogonal to the jet axis with the second segment resembling
a bow shock.  The first southeastern jet segment is narrower and 
brighter than the corresponding northwestern segment.  However, 
the area-integrated fluxes of the first jet segments
are about the same.  About 1\arcmin\ from the source, the jet exhibits an
abrupt bend towards PA = 10\arcdeg\ in the northwest region and towards
PA = 190\arcdeg\ in the southeast region of the jet. The portion of the jet 
beyond the  northwest bend is relatively brighter in [\sii ].
The spectra show no measurable Doppler shifts, in either \Ha\ or
[\sii ], although the line widths in \Ha\ are about 70 \kms .  The \Ha , [\nii ],
and [\sii ] lines do show a slight velocity gradient across the source star
with a roughly 20 \kms\ blueshift to the north and a comparable redshift
to the south.  These results imply that the jet lies very close to the 
plane of the sky with the northern beam pointed slightly towards us.

{\bf  HH 495:} 
This faint jet, located several arc minutes west of HH~7-11 and HH~12,
consists of 4 faint \Ha\ knots extending south from a faint star 
(Figure 12).

{\bf  HH 497:} 
This S-shaped flow (Figure~12) is centered on a bright star 3.2\arcmin\ east 
and 3.1\arcmin\ north of HH~366 at the eastern edge of NGC~1333.    
The southern side of the flow curves west and is bright in [\sii ] 
while the northern side curves east and is bright in
\Ha\ (Figure 13).  

{\bf  HH 498:} 
A curved filament of [\sii ] emission emerges from a bright
optical star about 5.5\arcmin\ due west of LkH$\alpha$~270
(Figure~9).  A very faint counter jet is also seen.  If this
feature is indeed a bent jet, it is the most extreme case found so 
far.  The bending angle is 40\arcdeg\ on the bright side of the jet and
about 35\arcdeg\ of the faint side.
The southern portion of the jet fell in the slit used to observe HH~334.
The radial velocity of the gas here is about +30 to +50 \kms .

{\bf  HH 499:}
This remarkable bent jet originates from an \Ha\ emission line 
star (EW[\ha ] = 138\AA ) located 3.1\arcmin\ west and 0.1\arcmin\ 
south of LkH$\alpha$~270, about 2.5\arcmin\  north of HH~12 (Figure~9).   
A nearly continuous, but bent 
filament of \Ha\ emission emerges from both sides of the star 
and ends in a knot of emission (HH~499 S1) 30\arcsec\ south of 
the star.  The filament is about twice as bright on the southern side 
of the star as on the north side.  On the north side of the star, 
the filament bends towards a knot located 35\arcsec\ northeast of 
the star (HH~499 NE1) and can be traced faintly to
a second knot, NE2, located another 25\arcsec\ away.   
The projected bending angle of the jet as measured from the star 
and original jet orientation is about 20\arcdeg . 

The spectra show that the southern (brighter) side of this jet 
is redshifted by about 50 \kms\ with faint wings extending up to
+100 \kms .    The fainter counter jet is hard to see in the
spectra, but appears to have a velocity of about $-$75 to $-$100
\kms\ with faint wings extending up to $-$125 \kms .
Both knots S1 and NE1 have predominantly redshifted emission
with velocities ranging from +25 to +50 \kms .

\section{Discussion}

It has become increasingly well documented that most stars are born 
in very high density and clustered environments.  
Both the Orion Nebula and NGC~1333 contain a few 
relatively massive stars which flood their environment   
with ultraviolet radiation.   The Trapezium cluster contains 
at least six stars with spectral types earlier than O9.5 which
produce a Lyman continuum ($\lambda <$ 912\AA ) luminosity of about   
L(LyC) = $2 \times 10^{49}$ photons~s$^{-1}$.  The NGC~1333
reflection nebula is illuminated by several late B type stars
with the most massive member, BD+30\arcdeg\ 549,  classified as B9.
Unlike the Orion Nebula, there is no \Hii\ region 
in this star forming region.  However, there is an intense
soft UV radiation field (912\AA\ $< \lambda <$ 2000\AA)
that can radiatively heat surrounding gas.

In externally irradiated outflow systems, the young stars powering  
outflows are frequently directly visible without significant 
obscuration.  Either these stars are older than the usual sources
of HH flows, or the surrounding cloud core has been 
prematurely eroded, possibly by the action of the external radiation 
field.

\subsection{Exposing the Sources; Photo-Evaporation of Cloud 
Cores and Envelopes} 

In a star forming environment containing O and early B type stars,
Lyman continuum radiation will ionize cloud surfaces and raise the 
temperature, pressure, and sound speed in the photon dominated layer.
Mass loss occurs from the cloud if the sound speed in this layer
exceeds the gravitational escape speed.  The Lyman continuum flux 
incident on a cloud surface is 
$F = L(LyC)/4 \pi d^2$, where $L(LyC)$ is the 
Lyman continuum luminosity and $d$ is the distance of the cloud
from the illuminating source. For a spherical cloud supporting a 
steady state outflow through a D-type ionization front at a radius $r$ 
the electron density at the base of the ionized flow is roughly
$n_e \approx (F / \alpha _B r )^{1/2}$ where $\alpha_B$ is the 
recombination coefficient.   This formula assumes that the outflowing 
plasma has spherical divergence, a roughly constant outflow speed, and
that the ionizing radiation field is absorbed in a shielding layer 
with a thickness comparable to the cloud radius.  It is assumed that 
mass loss occurs from the illuminated hemisphere only.   Since the 
flow leaves the ionization front at about the sound speed $c_{ii}$, 
the mass loss 
rate of freshly ionized plasma from the cloud surface is 
$\dot M = 2 \pi \mu m_H f c_{ii} (F / \alpha _B) ^{1/2} ~r^{3/2}$ 
where 
$\mu \approx 1.4$ is the mean molecular weight,
m$_H$ is the mass of hydrogen,  
$r$ is the ionization front radius, and 
$f$ is a number of order unity which depends on the
exact geometry and velocity profile of the accelerating flow 
(= 1/3 for a constant velocity spherical flow).   Thus,
$$
\dot M = 2.25 \times 10^{-5} {\rm (M_{\odot} yr^{-1})} f
         \biggl [ { L(LyC) \over 10^{49}}   \biggr ]^{1/2}
         \biggl [ { d \over {1~pc}}     \biggr ]^{-1}
         \biggl [ { r \over {0.1~pc}}   \biggr ]^{3/2}
         .
$$

However, even in the absence of ionizing 
radiation, the soft UV between $\lambda$ = 912\AA\ and about
2000\AA\ will heat the cloud surface layers.  At the base of the 
irradiated region, the heated layer can reach temperatures 
of 100 to over $10^3$ K, where the speed of sound $c_s$ is about 
1 to 3 \kms\  (cf. Hollenbach \& Tielens 1997, 1999; 
Johnstone, Hollenbach, \& Bally 1998).  Therefore,  soft-UV 
irradiated gas can expand from the heated surface 
so long as the local gravitational escape speed is less than 
the speed of sound, or where $r ~>~ 2GM(<r) / c_s^2$ (the so-called 
gravitational radius cf. Hollenbach \etal\ 1994; G is Newton's
gravitational constant and $M(<r)$ is the total mass inside radius 
$r$).   For a star plus circumstellar environment with a combined 
mass M (in Solar units), photo-erosion is likely to occur at 
distances larger than about 
1779 M $c_1^{-2}$ AU from the central star where the sound speed
at the irradiated surface, $c_1$, is in units of \kms .
Therefore, any residual envelope that may be left over from the 
formation of a low mass star will be dissipated
if UV radiation raises the sound speed at the cloud surface 
to more than the gravitational escape speed.

The time-scale for the evaporation of the residual envelope 
can be estimated by assuming that this soft UV radiation
is absorbed within a layer with a column density of 
$N_{21} = 10^{21}$ cm$^{-2}$, the penetration depth of this 
radiation (cf. Hollenbach \& Tielens 1997).  In the approximation 
that this flow diverges spherically, the base of the heated layer 
attains a density $n_{PDR} ~=~ fN_{21}/R$ where $R$ is the cloud 
radius.   If the actual gas density 
at radius $R$ is less than $n_{PDR}$,  the photon dominated region 
will advance into the cloud supersonically until a region with a 
density greater than $n_{PDR}$ is encountered.   However, in this 
higher density environment, the PDR heated front will advance 
sub-sonically; outside the transition from shielded to irradiated gas, 
a roughly steady state outflow is established.  In this scenario, 
the early transient stage is analogous to R-type ionization fronts 
while the later quasi-steady state is analogous to D-type ionization 
fronts.  If the gas density at the cloud surface is initially 
{\it larger} than $n_{PDR}$, the heating front will advance into 
the cloud sub-sonically.  Once steady-state flow is established, 
the density at the base of the flow will be given by $n_{PDR}$.  
The mass loss rate is given by $\dot M = 2 \pi \mu m_H c_s f N_{21} R$ 
so for typical parameters,   
$$
\dot M = 2.14 \times 10^{-5} {\rm (M_{\odot} yr^{-1})} f
         \biggl [ { N_{21} \over 10^{21}}   \biggr ]
         \biggl [ { r \over {0.1~pc}}       \biggr ] .
$$ 

When ionizing radiation {\it is} present, the softer UV penetrates
about a distance corresponding to $N_{21}$ 
into the cloud beyond the ionization front. 
Thus a PDR flow is established and may push the ionization front
to a somewhat larger radius determined by the conditions of 
photoionization equilibrium as shown by Johnstone \etal\ (1998), 

The approximate survival time-scale for an envelope surrounding a
young star is given by 
$$ 
\tau = {{M_e} \over {\dot M}} 
 = {{N_0 R} \over {3 f N_{21} c_s}}
$$
where $M_e$ is the total envelope mass and $N_0$ is the mean column density 
through this envelope.  For typical core parameters
($R$ = $r_{0.1pc}$ = 0.1 pc; $c_s$ = $c_3$ = 3 \kms ; 
$N_0$ = $N_{22}$ = 10$^{22}$ cm$^{-2}$),
$\dot M = 2 \times 10^{-5} r_{0.1pc} c_3 N_{22}$~M$_{\odot}$~yr$^{-1}$,
and 
$\tau = 2 \times 10^5 r_{0.1pc} c_3^{-1} N_{22}$ years.

Cloud cores and envelopes can often be represented by 
power-law radial density profiles of the
form $\rho (r) = \rho _0 r^{-\alpha}$.  It can be shown that
for mass loss controlled predominantly by the PDR, the cloud radius 
evolves with time according to
$$
r(t) = R \biggl [ 
  1 - \biggl ( {{2 - \alpha} \over { 3 - \alpha }} \biggr )
      {{ \beta } \over { \Gamma }}
      {{   t   } \over { R^{2 - \alpha}}}   \biggr ]^{1 / (2 - \alpha)}
$$ 
where $\beta = 2 \pi \mu m_H  f c_s N_{21}$ and 
$\Gamma = M_0 R^{\alpha - 3}$ and $M_0$ is the initial mass of the
power-law envelope inside radius $R$. 
As an example, for a power-law index of $\alpha = 1.5$, and an initial
envelope mass of 1 M$_{\odot}$ inside a radius of 0.1 pc, the envelope
is completely destroyed in $6.2 \times 10^4$ years for a PDR sound
speed of 3 \kms .    A uniform density cloud ($\alpha = 0$) with these 
parameters evaporates completely in $7 \times 10^4$ years.  
Thus, envelope evaporation is rapid and insensitive to the details
of the cloud structure.

For mass loss controlled predominantly by Lyman continuum radiation, 
$$
r(t) = R \biggl [ 
  1 - \biggl ( {{3 - 2 \alpha} \over { 2 (3 - \alpha) }} \biggr )
      {{ \gamma } \over { \Gamma }}
      {{   t   } \over { R^{(3 - 3 \alpha)/2}}}   \biggr ]^{2 / (3 - 2 \alpha)}
$$ 
where $\gamma = \mu m_H f c_{ii} ( \pi L(LyC) / \alpha_B )^{1/2} d^{-1}$ 
and $d$ is the distance from the source of ionization.   For
the parameters used above, complete envelope ionization is somewhat
slower, taking about $3 \times 10^5$ years for the uniform density case.

While the dissipating envelopes surrounding young stars in NGC~1333
are expected to remain predominantly neutral,  within the Orion Nebula
and its photo-ionized environment, the expanding envelopes will pass
though an ionization front to form bright expanding condensations
within the body of the \Hii\ region.  In the Orion Nebula proplyds, 
mass loss is sustained by the continued photo-evaporation of dense 
circumstellar disks at mass loss rates of about $\dot M = 10^{-7}$   
M$_{\odot}$~yr$^{-1}$ (Johnstone, Hollenbach, \& Bally 1998).  Since 
these disks are much smaller than the cloud cores considered above 
($\sim$100 AU vs. 0.1 pc), their mass loss rates are correspondingly 
much lower.  {\it The heating, expansion, and removal of lower density 
material will decrease the extinction and render visible the immediate 
circumstellar environment, including the young stars, their jets, and winds.}

Dissociation,  heating, and photo-evaporation by the soft UV 
radiation fields of nearby massive O, B, and even early A stars  
may play a major role in destroying the envelopes from which young stars 
accrete, thereby bringing further stellar growth to a halt.  
Since irradiated jets appear to be associated with naked young stars,
their presence may signal one method by which Nature regulates the 
final masses of stars, namely the termination of accretion by external 
photo-evaporation.  Furthermore, the photo-evaporative flows may
play a fundamental role in the origins of turbulent motions inside
molecular clouds (Bally \etal\ 2000a).

\subsection{The Utility of Irradiated Jets}

The launching of a high velocity wind or jet requires a reservoir of 
mass in the circumstellar environment to fuel the outflow.  Most
likely, this reservoir is in the form of a compact high density
circumstellar disk surrounding the star.  The sources of the 
irradiated jets must be in an evolutionary state where they have lost 
their envelopes (since the stars suffer low extinction) but still retain 
a substantial disk (since they power an outflow).

When such an outflow or jet is irradiated from the outside, the
radiation field can dramatically improve the visibility of the moving
fluid both in and far from the shock zone.   Therefore, as noted by
Reipurth \etal\ (1998),
Herbig-Haro flows embedded within \Hii\ regions provide a unique 
opportunity to determine completely the physical parameters of an 
outflow.  Ionizing radiation in such an environment can partially or 
fully ionize all or most outflow components.  In this situation, 
shock waves are {\it not} required to render the gas visible.  
Raga \etal\ (2000) present a more complete discussion of the
propagation of ionization fronts into externally irradiated 
cylindrical jets. 

Even when Lyman continuum radiation is not present, the softer
UV radiation can enhance the visibility of the outflow.
Photo-electric heating resulting from the interaction of the soft
UV radiation field with grains in the medium produce extended  
regions that are bright in the optical forbidden lines such as
[\sii ].

\subsection{Photo-Ionization of Irradiated Jets}

Assume that the jet is launched
with a steady ejection velocity, $v_j$ and prior to being
irradiated can be characterized
by a constant sound speed along its length, $c_j$.  These 
assumptions imply that the jet, if permitted by the ambient medium,
will spread orthogonal to its    
direction of propagation at the Mach angle, $\theta _j = 2 c_j / v_j$
so that the radius of the jet, $r_j$, as a function of distance
from the source, $d$, increases as $r_j = ( c_j / v_j ) d$.
If the mass loss rate into this lobe of the jet
is given by $\dot M_j$, the mean density along the 
jet decreases as 

$$
n_j(d) ~=~ \dot M_j / ( \pi \mu m_H r_j^2 v_j)
~ = ~ \dot M_j v_j / ( \pi \mu m_H c_j^2 d^2)
$$   

Now consider a source of Lyman continuum radiation illuminating
this jet so that the angle between the jet outflow axis and the
direction to the illuminating star is given by an angle $\theta$.
If the jet points away from the ionizing source, it will initially be
shadowed by the circumstellar disk.
Once the jet emerges from this shadow the incident flux of 
Lyman continuum radiation is given by 

$$
F ~=~ L(LyC) sin ( \theta ) e^{- \tau (UV)} / 4 \pi D^2
$$

\noindent
where $\tau (UV)$ is the total opacity of the medium between
the source of Lyman continuum radiation 
which is assumed to be at a distance $D$ from the portion of the
jet under consideration. 

Under most conditions, an R-type front will propagate rapidly into 
the side of the jet facing the illuminating star.    However, Raga \etal\
(2000) show that under some conditions the front can also start out as a
D-type front.   If the jet
has a sufficiently low density, it will become fully ionized by 
the incident radiation field.   On the other hand, for a sufficiently    
high density jet,  recombinations in the photo-ionized layer will
slow the advance of the R-type front, eventually converting it    
into a D-type front.  However, the gradual divergence of 
the jet guarantees that even in this latter case, as the distance from 
the jet source increases, an ever increasing portion of the jet 
will become ionized until the whole jet is fully ionized.  
In either case the degree of penetration of the ionizing radiation 
depends on the azimuthal angle about the jet axis.  Penetration is 
greatest on the side of the jet facing the ionizing source, 
and least on the side facing away from the source.    Back-side 
ionization is produced by the diffuse radiation field present 
in the \Hii\ region and is therefore sensitive to the detailed spatial 
and density structure of the \Hii\ region.   

Ionization will raise the plasma temperature of the jet to about 6,000 to 10,000 K 
for normal (Solar) abundances.  Because the sound speed in the neutral 
shielded portion of the jet is likely to be less than the sound speed 
in the ionized plasma, photo-ionization will likely increase the divergence 
(Mach) angle of the jet.  In the ionized region, the transverse expansion 
orthogonal to the jet axis will occur at about a velocity of 
$c_{ii}$, the sound speed in the photo-ionized medium. 

The incident Lyman continuum radiation is completely consumed by 
recombining ions and electrons in a layer of plasma with a thickness
$h = F / \alpha _B n_j(d)^2$ where $n_j(d)$ is the electron density
at the ionization front.  If at the point where the jet emerges into 
the irradiated zone, $r_j  <~ h$, then the jet is in the low density 
limit and it will become fully ionized as the R-type front races through 
the thickness of the jet.  If  $r_j  >  h$, the jet is in the high density 
limit and full ionization will be regulated by the divergence of the jet 
at the Mach angle.
 
For a dense jet, setting $r_j  =  h$ leads to a an estimate of the 
maximum distance from the source of the jet at which a neutral core can 
exist, $r_{max}$.  Beyond this distance, the jet will be fully ionized.  
Therefore,

$$
r_{max} ~\approx ~ 
{{2 \alpha _B \dot M_j v_j}
   \over
 {\pi ^2 \mu ^2 m_H^2 c_j^3 F }} 
$$
This assumes that the radiation field is symmetric about
the jet axis and that the core is ionized at the same rate in all directions.
In general, the radiation field will be much stronger on one side of the jet
and the shadowed side will become ionized more slowly.  Highly asymmetric
ionization can increase $r_{max}$ by about a factor of 2.
For symmetric illumination,  $L(LyC) ~=~ 10^{48}$, $v_j$ = 200 \kms , 
$c_j$ = 10 \kms , $D$ = 3 pc, $\dot M$ = $10^{-8}$ \msol $\rm yr^{-1}$, and
$sin( \theta ) ~=$ 0.5,  we find $r_{max} ~=~5.8 \times 10^{16}$ cm
and the density of the jet at $r_{max}$ is  
$n_j(r_{max}) ~=~555 \rm {cm^{-3}}$. 
This corresponds to about 4000 AU or 0.02 pc, which is short
compared to most HH jets.

Most Herbig-Haro flows are episodic.  The shock waves associated
with internal working surfaces, formed where faster ejecta overtake 
slower moving material,  make portions of the jet visible.   
The four irradiated jets investigated by Reipurth \etal\ (1998) 
in the $\sigma$ Orionis region exhibit relatively strong
[\oi ] and [\sii ] emission (compared to \Ha ) close to the source
where the jet is either shielded from external ultraviolet radiation or
still retains a neutral core.  This forbidden line emission
originates in compact knots which likely trace internal working
surfaces within the body of the jet.  The brightest $\sigma$ Orionis jet, 
HH~444, is associated with several nested bow shocks located more than 
1 arc minute from the source. These features provide evidence that 
irradiated jets are also likely to be powered by 
time variable mass loss.

The episodic mass loss produces jets with density variations along their 
lengths. Furthermore, internal working surfaces and complex shock structures
will form where faster ejecta overtakes slower ejecta supersonically.
Internal working surfaces can produce line emission in the shielded
portion of the jet.  When low excitation shocks dominate excitation,
the resulting emission will be dominated by the forbidden emission lines
of neutral and low ionization states of common species like [\oi ],
[\sii ], and [\nii ].   As such a jet emerges into the external radiation
field, the outer layer of the outflow will become ionized and the interior
will be heated by the FUV radiation field.  Thus, the jet skin will develop
an \hii\ region spectrum while the shielded jet interior will exhibit enhanced
forbidden line emission due to the heating by the FUV radiation.
However, divergence of the flow insures that eventually, the entire jet
becomes photoionized and will therefore produce an \hii\ region spectrum.
Internal working surfaces within the jet will alter the density distribution.
High density regions may be self-shielded and may retain mostly neutral cores
for a much longer time than the lower density regions.  Thus, episodic
behavior and variations in the jet ejection velocity will generate 
clumps and complex structure in the irradiated portion of the jet.  However,
external ionization of these structures will render all moving fluid 
elements visible on deep images.

\subsection{Asymmetries in Irradiated Jets}

The four irradiated jets discovered by Reipurth \etal\ (1998) 
in the $\sigma$ Orionis sub-group of the Orion OB association
are predominantly one sided.  The brightest portions of these
outflows point {\it away} from the hot star.
Preliminary spectroscopic evidence (cf. Bally \etal\ 2000b)
indicates that for the jets near $\sigma$ Ori, the brighter jet beams
often have lower speeds than fainter counter jets.  Therefore,
the apparent asymmetry in jet intensity may be a consequence of
an underlying {\it kinematic asymmetry}, possibly produced by the
radiation field.  The radiation field may
preferentially decrease the density of matter on the irradiated side
of a disk more than on the shadowed side.  As an otherwise symmetric
jet (both beams have the same mass loss rates and speeds)
propagates through such an asymmetric medium, it may suffer more
mass loading on the denser shadowed side than on the less dense
irradiated side.   This is especially true if mass loading primarily 
effects the skin of the jet, as it is the skin that is first ionized.
Such mass loading could decelerate the jet motion through the shadowed 
side more than on the irradiated side.

Bally \etal\ (2000a) discuss over 20 irradiated microjets embedded within the
core of the Orion Nebula which were discovered on deep high resolution
Hubble Space Telescope images.  The majority of these jets are also
asymmetric with a slight tendency for the brighter beam to face the 
dominant source of the UV radiation.
Hirth \etal\ (1997) found that many microjets produced by
T-Tauri stars appear to be one sided; some show a kinematic asymmetry
with one beam having a larger velocity.

All five outflows reported here in the outskirts
of the Orion Nebula, HH~502 -- 506, are also asymmetric.  
Although both HH~502 and 503 
contain bow shocks on either side of the source, indicating that mass loss
is bipolar, the jets in both of these sources are more prominent on the side
of the source facing away from the source of Lyman continuum radiation.
Though jets are visible on the opposite side 
of the source that faces into the radiation field, they tend to be much fainter
and shorter.   

As discussed by Reipurth \etal\ (1998), extinction can be ruled out
as an explanation for the asymmetry of the $\sigma$ Orionis jets.   Similar
arguments can be used to rule out extinction as the explanation for the
asymmetries of the jets located in the Orion Nebula.  The nebular
background is sufficiently bright so that any extended absorbing cloud would
be seen in silhouette.  Furthermore, in the two jets where bow shocks are seen
on both sides of the central source, these shocks have similar surface 
brightness, indicating that there is little difference in extinction on 
the two sides of the source.  

The preliminary evidence suggests that the Orion jets exposed to Lyman
continuum radiation are often one sided.  On the other hand, the irradiated
jets in NGC~1333, which are exposed to weaker and softer UV radiation fields,
tend to exhibit bipolar symmetry.

\subsection{The Origins of Jet/Counterjet Brightness Asymmetry 
in Irradiated Outflows}

In this section, two models of bipolar jets which can produce the 
observed asymmetries in irradiated jets are discussed.  It is assumed 
that the jet is intrinsically bipolar with the same mass loss rate, 
$\dot M_j$, into each beam.   However, the velocity with which each 
beam moves away from the source, $v_j$, is assumed to be different for 
each beam.   As discussed above, different amounts of mass loading in 
each beam may produce such kinematic asymmetry. 

In the first model, case A, the opening angles of the lobes of the 
bipolar jet are assumed to be the same.   Lobes of a bipolar jet may 
expand into cones having the same opening angle if jet spreading is 
dominated by {\it external processes} such as a high pressure external medium 
or one that contains a strong magnetic field.   In kinematically asymmetric 
bipolar jets with a {\it constant beam opening angle} the {\it slower} 
beam is brighter.   In the second model,  case B, 
the speed at which jet material spreads orthogonal to its bulk motion 
is assumed to be constant.  Therefore, the opening angle 
of the jet is inversely proportional to the jet speed.  This situation is 
likely to prevail if the transverse expansion of the jet is determined by 
{\it internal processes} such as heating by a radiation field so 
that each beam expands at the sound speed into a Mach cone.   
Alternatively, if the amplitudes of the velocity variations producing 
internal shocks are the same in each beam,   splashing of jet material 
orthogonal to the jet axis may occur at a fixed velocity.  
In kinematically asymmetric bipolar jets with a {\it constant transverse 
spreading}, the {\it faster} beam is brighter.  

Other, more complex models, which will not be discussed here in
detail, are also possible.  For example, if the kinematic asymmetry is
generated at the source rather than by mass loading, the faster 
outflow lobe might also have time-variability with a larger velocity 
amplitude.   This would produce stronger shocks, and denser shocked 
material, which would result in a brighter jet beam.

\bigskip
\centerline{\it Case A: Bipolar Jets With a Constant Opening Angle}

Consider a bipolar jet in which both lobes have the {\it same} opening angle,
$\theta_j$.  Furthermore, assume that the mass loss rates into each beam are 
identical, and that the beams have different velocities.  Therefore, 
at a given distance from the source $d$,  the jet radii, 
$r_j(d) = \theta_j d / 2$ will be the same.  However, the continuity 
equation implies that if the mass loss rates, $\dot M_j$, into each beam 
are the same, the density of matter in the jet, 
$n_j(d) = \dot M / ( \pi \mu m_H r^2_j(d) v_j $,  
will be higher in the slower beam than the corresponding density at the 
same distance from the source in the faster beam.

The surface brightness of an emission line scales as the emission measure, 
$EM = n_e^2 l$, where $l$ is the path length of the observer's line-of-sight 
through the emission region.   Therefore, for a fully ionized bipolar jet 
which has {\it the same opening angle} in each beam, the faster beam 
will appear fainter by a factor $(v_{fast}/v_{slow})^2$.    In general, 
$$
EM \propto  \biggl [ {{ \dot M } \over { \pi \mu m_H  }} \biggr ] ^2
            {{ 2 } \over { \theta ^3_j d^3 v_j^2 }}. 
$$ 
When the contrast between the \hii\ region and the jet is low, the
fainter jet can altogether vanish from view.

\bigskip
\centerline{\it Case B: Bipolar Jets With Constant Transverse Spread}

On the other hand, if the jet lateral spreading occurs at a constant
transverse velocity, $v_t$,
each beam of the jet will exhibit a similar density and emission 
measure at a comparable dynamical age as opposed to at a comparable 
distance from the source.   Because the opening angle is given by 
$\theta _j = v_t / v_j$, the faster beam will have a narrower opening 
angle and higher density at a given distance from the source.
Using this definition of $\theta _j$ in the formula above, the emission
measure is found to scale as 
$$   
EM \propto  \biggl [ {{ \dot M } \over { \pi \mu m_H }} \biggr ] ^2
            {{ 2 v_j } \over { v_t^3  d^3 }}  .
$$
Thus, for the case where the opening angle is defined by a constant
transverse velocity, the
{\it faster} beam will exhibit the greater surface brightness
at a given distance, a prediction opposite to that of a  
constant opening angle jet.   

In summary, if external processes dominate jet expansion, and the opening 
angles of each lobe are the same, then at a given distance 
from the source, each beam will have the same transverse extent (Case A). The 
slower beam will then be denser and brighter. Alternatively, if internal 
processes such as splashing from internal working surfaces or heating 
by a radiation field dominate the transverse spreading of the jet, 
each lobe is likely to expand at the same transverse velocity (Case B).  
When the jet speeds are not the same, the slower beam will exhibit a wider 
opening angle.  Therefore, at a given distance from the source, the faster 
beam will be narrower, denser, and brighter.  High dispersion spectroscopy 
and imaging will be able to distinguish between these different scenarios 
and probe the generality and origins of kinematic asymmetry.

\subsection{Jet Density Estimation for Irradiated Jets}

The \Ha\ surface brightness can be used to estimate the emission 
measure, EM, of a jet from the relationship 
$EM(\Ha ) = 4.89 \times 10^{17} I(\Ha )$ $(cm^{-6} pc)$ where
$I(\Ha )$ is in units of $\rm erg s^{-1} cm^{-2} arcsec^{-2}$
(Spitzer 1978).  If the extent of the jet along the line-of-sight 
is assumed to be the same as the projected width of a feature, 
the emission measure can be used to estimate the electron density of 
the plasma within the jet from the relation 
$n_e = ( EM / l(pc) )^{0.5}$ where $l(pc)$ is the path length 
through the emission feature.   These quantities are tabulated in
Table~3 along with the surface brightness of the [\sii ] emission
from the same features and the \Ha\ to [\sii ] surface
brightness ratio for a selected set of points in the major detected
features of each flow.  The listed values correspond to the
peak surface brightness in each region, and therefore yield upper bounds
on the line intensities and derived electron densities.     

If the jet is fully ionized by the external radiation field
the derived plasma density {\it is} the jet density, which can be combined
with measurements of the jet velocity to estimate the mass loss rates into
each beam, the momentum, and energy fluxes.
The flow momentum and energy can be estimated from the density, 
the flow cross-section, and the fluid velocity.   
For a cylindrical flow with radius $r_j$, density $n_j$, and
velocity $v_j$, the mass loss rate in the jet, or equivalently
the rate at which mass flows through a given stationary surface, is 
given by $\dot M = \pi \mu m_H n_j v_j r_j^2$.  In terms of this 
quantity, the flux of momentum and kinetic energy are
$\dot P = \dot M v_j$ and $\dot E = 0.5 \dot M v_j^2$.  Thus, for
a jet speed of $v_{100}$ in units of 100 \kms , a
jet density $n_3$ in units of $10^3$ $cm^{-3}$, and
a jet radius $r_{115}$ in units of 115 AU (which corresponds to
our seeing deconvolved resolution limit in Orion),  
the jet has a mass loss rate 
$\dot M = 3.4 \times 10^{-9} v_{100} n_3 r_{115}^2$ M$_{\odot} yr^{-1}$,
a momentum transfer rate of 
$\dot P = 3.4 \times 10^{-7}v_{100}^2 n_3 r_{115}^2$ 
M$_{\odot} yr^{-1} km s^{-1}$, and
a mechanical luminosity of 
$2.7 \times 10^{-3} v_{100}^3 n_3 r_{115}^2$ L$_{\odot}$. 
Table 4 summarizes the mechanical parameters of the photo-ionized
portions of the flows in the Orion Nebula. 

The extent to which the jet is ionized can be estimated by
comparing the observed emission measure to that of a self-shielded 
cylinder of neutral gas.   If the jet contains a shielded neutral
core with a radius $r_n$, and is located at a distance $D$
from a source of Lyman continuum radiation with luminosity
$L(LyC)$, then photo-ionization equilibrium in the shielding layer implies that
$$
{{L(LyC)}  \over {4 \pi D^2}} = a n_e^2 \alpha _B r_n
$$
where $a$ is a factor of order unity which takes into account the
relative orientation of the jet,  the line of sight to the illuminating
star,  and the geometric divergence of the expanding plasma.  Thus, the
emission measure is given by
$$
EM = {L(LyC) \over {4 \pi a \alpha _B D^2}} = 
2 \times 10^5 D_{pc}^2 ~~~~~~~~~~~(cm^{-6} pc)
$$ 
where $L(LyC) = 2 \times 10^{49}$ photons per second is 
the Lyman continuum luminosity of the Trapezium,  and $D_{pc}$ is in
parsecs.  The inner portion 
of the HH 502 and 503 jets and some of the brighter knots in the HH~502
bow shocks are only a factor of 4 -- 10 lower than this maximal 
emission measure.   Thus, these portions of
the flow are likely to be only partially ionized with a shielding layer
that protects a neutral core from the Lyman continuum radiation.  
Thus, the mass loss estimates and other parameters listed in Table 5
may well be lower limits.
  
On the other hand, finite angular resolution may result in an
overestimate of the jet mass loss rate.  Near their sources, the Orion
Nebula jets are essentially unresolved in width.  Thus, the jet width 
used in the estimation of mass loss rates is {\it assumed} to be at 
the seeing-deconvolved resolution limit of about 0.5\arcsec , corresponding
to a jet radius of 115 AU.   If the jet radius is actually {\it smaller}, 
the estimated electron density in the ionized region {\it increases} 
as $r_j^{-0.5}$, and the resulting jet mass loss rate {\it declines}
as $r_j^{1.5}$.  Thus, higher angular resolution measurements of both
the jet ionization structure and its width, and a combination of
proper motion measurements and more precise radial velocity determinations
are needed to better constrain the jet parameters and to take full
advantage of the ionized nature of these outflows.

\subsection{Why Are Irradiated Jets Highly Collimated? }

HH~502 in Orion (Figure 2) and HH~333 in Perseus (Figure 8) are the longest
jets discussed here.  And they are both extremely collimated.
The beams of these jets remain less than two arc seconds in diameter
at distances larger than 40\arcsec\ (for HH~502) and 120\arcsec\ (for HH~333)
from their respective sources.  Thus, these jets remain collimated to within
a few degrees or less.   Such a high degree of collimation is even more
surprising in the light of the low radial velocities measured for these jets.

In an externally photo-ionized jet such as HH~502 or HH~503, the sound speed
in the jet interior should be about 10 \kms .  If the intrinsic jet speed is
assumed to be 100 \kms , then the Mach angle is predicted to be about 6\arcdeg ,
and the jet opening angle should be about 12\arcdeg ,  many times larger than the
apparent opening angles of HH~502 or HH~503.  There are three possible reasons
why these jets may appear so collimated.

First, these jets may lie nearly in the plane of the sky, in which case
their true velocities may be many times the observed radial velocity.  However,
to produce a less than 3\arcdeg\ opening angle, the speed of the HH~502 or
HH~503 jets would have to exceed 300 \kms .

Second, it is possible that the jet is so dense that only a thin surface layer
becomes ionized throughout its visible length.   Thus, the jet density estimates 
based on the emission measure may be strict lower limits.  In this scenario, the jet
interior may remain sufficiently shielded and cold so that the Mach angle remains 
very small and the jet does not spread.  However, to explain the fading or 
disappearance of the beam, the beam must come to an end at some distance from 
the source.  A time-variable outflow would produce segments of ejecta separated 
by relatively empty gaps.   Perhaps the apparent ends of these jets
correspond to the beginning of the gaps produced by lower mass loss rates 
in the past.   The bow shocks seen downstream of the apparent ends of jets 
such as HH~333 and HH~502 may trace the debris piled up by earlier high mass loss 
rate episodes that were followed by relative inactivity.  This explanation is 
most likely to apply to the highly collimated irradiated jets in the Orion Nebula.

Third, there may be an invisible confining medium that prevents free expansion.
Such a medium may consist of a dense but invisible gas, or a magnetic field.
However, a dense confining medium in the Orion Nebula's interior would be
ionized and visible.  So if confinement is invoked for Orion's irradiated
jets, then a magnetic field of sufficient strength must be present.  However,
in NGC~1333, a dense medium could remain invisible.  But, its total dust content
must not be sufficient to hide the jet. Additional observations are needed to
decide which model is the best explanation for the high degree of collimation
of the irradiated jets.

\subsection{Origin of the C-Shaped Symmetry of Jets in H II Regions}

Two distinct types of C-shape symmetry are observed in Orion and
NGC~1333.  All of the bent jets embedded within the Orion nebula 
bend {\it away} from the nebular core.  On the other hand, the 
bent jets in the core of NGC~1333 tend to bend {\it toward} the 
cluster core.

There are three forces which can in principle deflect a jet propagating
through the ionized interior of the Orion Nebula;  ram pressure
of the champagne flow of plasma from the nebular interior ($P_{ram}$),
radiation pressure ($P_{rad}$), and the rocket effect of the jet shielding
layer pushing on the jet neutral core ($P_{rock}$).   

Evidence for a general, widespread outflow from the core of the
Orion Nebula has been provided by high dispersion
spectroscopy (O'Dell 1994; Wen \& O'Dell 1993; O'Dell \etal\ 1993) 
and by the presence of stationary
bow shocks formed where T Tauri star winds impact this outflow
(Bally \etal\ 2000a).  Though highly uncertain, the
mass flux from the Orion Nebula core has been estimated to be
in the range $\dot M$ = $10^{-4}$ to $10^{-3}$ M$_{\odot}$~yr$^{-1}$
with a speed of order 20 \kms .  Bally \etal\ (2000a) interpreted the
so-called LL Orionis wind-wind collision fronts in terms of a collision 
between wide-angle T Tauri stellar winds and the champagne flow 
from the Nebula.  Table 5 lists additional wind-wind collision 
fronts located in the vicinity of the irradiated jets which provide 
evidence for such a bulk outflow of plasma from the Nebula.
Although stellar winds from the hot Trapezium stars could also 
play a role, the $\dot M V$ product of the champagne flow is likely 
to exceed that of the massive star winds by more than an order of 
magnitude in the outer nebula.  The effect of the champagne flow 
ram pressure is identical to the case of a jet propagating in 
a cross wind (cf. Raga \etal\ 1995).

The champagne flow ram pressure can be estimated from the mean density 
of the nebula near the bent jets, and its average flow velocity 
of slightly more than the sound speed in ionized gas.  Thus, 
$P_{ram} = \mu m_H n_{ii} (f c_{ii})^2$ 
$\approx 1.4 \times 10^{-10} n_{100} c_{ii}^2$ dynes cm$^{-2}$ where
$n_{100}$ is the plasma density in units of $\rm cm^{-3}$ and $f$ is a factor
of order unity.  Radiation pressure is given by 
$P_{rad} = \tau L / 4 \pi d^2 c$ 
$\approx 1.1 \times 10^{-10} \tau L_5 d_{pc}^{-2}$ dynes cm$^{-2}$ where
$L_5$ is the luminosity in units of $10^5$ L$_{\odot}$, $\tau$ is the
opacity of the medium to incident radiation (most likely much less than 1),  
and $d_{pc}$ is the distance from the light source in parsecs.
Finally, for jets above the critical density, an ionized layer
forms on the illuminated side of the jet.  Its expansion pushes
on the neutral jet core, deflecting it away from the light source by
means of the rocket effect.   The resulting pressure is given by
$P_{rock} = \mu m_H n_j (f c_{ii})^2$ 
$\approx 1.4 \times 10^{-9} n_{1000} c_{ii}^2$ where $n_j$ is the observed
plasma density in the ionized portion of the jet (cf. Table 3).     
Typically, the jet surface layer is much more than an order
of magnitude denser than the \hii\ region density in the vicinity of
the jet.  If this were not the case, the jet would not be visible 
against the background nebula since the path length through the jet 
is usually smaller than the path length through the nebula by more 
than three orders of magnitude.  Once the jet becomes fully ionized
(due to spreading at the internal Mach angle), the rocket effect 
disappears since the fully ionized jet plasma will expand symmetrically 
about its axis.   The rocket effect is most effective in bending the
jet when about half of the beam is ionized and the amount of bending
expected is roughly $\theta = c_{ii} / v_j$ which is just the Mach angle 
for a fully ionized jet.  For the parameters appropriate for HH~502 to 505,  
$P_{rock} > P_{ram} \approx P_{rad}$.  And the rocket effect is likely
to dominate other external forces which can act directly on the jet beam.

However, it is possible that deflection of the jet cocoon also plays a 
role in bending these outflows away from the nebular core.
The densest parts of these cocoons are traced by the various bow shocks
associated with the Orion Nebula irradiated jets.   However, these
features are likely to be fully ionized and optically thin to incident
radiation.  Therefore, the only force capable of deflecting them is 
the ram pressure of the expanding nebula.
Although capable of deflecting the cocoon walls downstream, the cocoon
interior is shielded from this force.  Such deflection of the cocoon
may be responsible for making the upstream side of the HH~503 W1 bow shock
be so close to the HH~503 jet.  It also may be responsible for the LL 
Orionis type bow shock surrounding HH~505.   However, if this was the 
only force acting on the system, the jet should propagate from the source 
without deflection until it impacts the deflected cocoon wall.   But, our 
images of HH~502 and 503 show that the jets themselves are deflected.   
If the cocoon interior were sufficiently magnetized, it is possible 
that Alfv\'en waves in the cocoon interior could transmit the pressure 
applied on the cocoon walls to the jet, causing it to bend.   However, 
it is more likely that the above mentioned rocket effect deflects the 
jet while the ram pressure of the champagne flow deflects the jet 
cocoon and the associated bow shocks.  The observed deflection angles of 
both the jets and the jet cavity walls can be used to confront these 
theoretical ideas. As shown in Table 2A,  the observed jet bending angles 
are of order 10\arcdeg .  

The true bending angle $\theta$ of a jet may be quite different from the 
observed bending angle due to the projection of the 
outflow onto the plane of the sky.    Consider a bent outflow from a 
source assumed to lie at the origin of a Cartesian coordinate system.  
Assume that one beam of the outflow propagates in the $-x$ direction.  
This jet and its counterjet define a plane in which the jet bending
angle can be defined as the angle $\theta$ between a vector directed 
along the $x$ axis (along the expected direction of the counterjet in the
absence of bending) and the counter-jet. The y-axis of the coordinate 
system is assumed to lie in the plane of the curved jet.   The projection of 
the outflow onto the plane of the sky is described by two rotation 
angles,  $\alpha_x$ and $\alpha_y$, about the $x$ and $y$ axes, respectively.  
(In general, an additional rotation about the line-of-sight is also needed 
to align the jet or counter-jet with a specific orientation on the sky, 
but this rotation does not effect the projected jet deflection angle.)  
In terms of these two angles, the observed (projected) deflection angle, 
$\theta _P$ is given by
$$
\theta _P ~=~ tan^{-1}
\biggl [
  {{  tan (\theta)  cos (\alpha_x) } 
   \over
   {  cos (\alpha_y) ~-~ tan (\theta) sin (\alpha_x) sin (\alpha_y) }}
\biggr ]
$$

For example, if $\alpha_x$ = 30\arcdeg\ and $\alpha_y$ = 60\arcdeg\
the true bending angle is nearly a factor of two smaller than the 
observed (projected) bending angle. However, the inclination angles are 
unknown for the jets discussed here.  

Cant\'o \& Raga (1995) discuss the deflection of a jet by a supersonic 
side wind.   In this model, the ambient medium interacts directly 
with the  jet rather than a surrounding cocoon formed by sideways 
splashing of material colliding in internal working surfaces.  The 
jet deflection angle, $\alpha$, is defined as the angle between the 
launch orientation and a line tangent to the jet at some point.  
For an isothermal jet with sound speed $c_j$, this angle is given by
$$
\alpha  ~=~ tan^{-1} 
\biggl [
   \biggl ( {{ \rho_a } \over {\rho _j}} \biggr )^{1/2} 
   \biggl ( {{ x_j } \over { r_j }} \biggr ) 
   \biggl ( {{ v_a } \over { v_j }} \biggr ) 
   \biggl ( {{ 1   } \over { M_j }} \biggr ) 
\biggr ]
$$ 
where 
$\rho_a$ and $\rho _j$ are the densities of the ambient moving medium 
and the jet, 
$v_a$ and $v_j$ are the velocities of the ambient medium and the jet 
(assumed to be initially perpendicular), 
$x_j$ is the distance from the jet source of the point where the jet 
bending is being measured, 
$r_j$ is the radius of the jet at that point, 
and $M_j = v_j / c_j$ is the Mach number of the jet.
Given an observed deflection angle, this formula can be solved for
$v_a$, the relative motion between the jet source and the medium.
For small angles, $\alpha$ = 2 $\theta_P$ so that for typical 
inclination angles, the observed deflection angles (Table 5) can be 
directly used in the above formula.  
For $v_j = 100$ \kms , $\alpha= 10$\arcdeg , 
$r_j/x_j$ = 1/30, 
$$
v_a = 5.3 \biggl ( {{ \rho_j } \over {\rho _a}} \biggr )^{1/2} ~~~~~~
( \rm km ~s^{-1}  )
$$
Since the Orion jets are visible against the nebular background, they
must be much denser than the surrounding medium.  
For $n_j = 10^3$~cm$^{-3}$
(Table 3) and $n_a = 100$~cm$^{-3}$ (a reasonable guess for the 
outskirts of the Orion Nebula),  $v_a \sim 17$ \kms.  

This velocity is about what is expected for either deflection
of the jet by the rocket effect, or by the ram pressure of 
nebular champagne flow.  If the ambient medium is interacting with the 
jet cocoon rather than the jet itself (as may be the case for HH~502 and 
503), the density $\rho_j$ above should be that of the cocoon, which may 
be comparable to the ambient medium.  In this case a considerably lower 
velocity side wind can deflect the flow.
Thus, {\it the C-shaped symmetry of the Orion Nebula irradiated jets
may be produced by the combined action of the rocket effect acting on
the ionized skins of the jets, and the ram pressure of the nebular
champagne flow acting on the outer walls of the cavities created by the
jets (the jet cocoons).} 

\subsection{Origin of the C-Shaped Symmetry in Dense Clusters}

There are two important differences between the jets embedded within
NGC~1333 and those in the Orion Nebula.  First, all bent jets 
in the Orion Nebula appear to {\it lie at large projected distances from 
the cluster core}.  Most lie several parsecs from the Trapezium cluster,
well beyond its dynamical influence.  At this distance, the time scales
for crossing the cluster at the expected stellar velocities  
are much longer than the ages of the observed HH flow components. 
In contrast,  all of the bent jets in the NGC~1333 region {\it lie within
the projected boundaries of the embedded cluster}. 
Second, the bent jets within the Orion Nebula are exposed to 
outward oriented forces such as the rocket effect and the 
low Mach number outflow of plasma from the nebula core.
In contrast, {\it there is no rocket effect or outflow of plasma
from the NGC~1333 core since there is no H~II region}.
The jets in NGC~1333 are only irradiated by non-ionizing soft-UV
photons.  This difference eliminates the bending mechanisms 
proposed for the irradiated jets in the Orion Nebula.   Furthermore, 
it makes it conceivable that the bent jets in NGC~1333 result 
from recent dynamical interactions in the rich cluster of 150 stars 
packed into a region smaller than 1 pc in diameter.

C-shaped outflows can be produced when either the medium moves past the
source of a bipolar jet, or the source moves through the medium.  In
the Orion Nebula, the former explanation plus the outward push of the
rocket effect on the jet skin can explain the jet bending.  However, 
the bipolar flows HH 334, HH~499, and HH~498 near NGC~1333 bend 
{\it towards} the core of the star cluster.    For intrinsic bending angles
comparable to the observed values, 
HH~334 requires a transverse velocity of about 
$2.9 (\rho_j/\rho_a)^{1/2} v_{100}^2$ \kms .  
However, HH~499 and HH~498 require transverse motions of 
$14 (\rho_j/\rho_a)^{1/2} v_{100}^2$ \kms\ and 
$40 (\rho_j/\rho_a)^{1/2} v_{100}^2$  \kms , respectively.   
It is likely that HH~498 and 499 are seen nearly
end-on so that projection effects greatly exaggerate the bending angle.
Even so, these velocities are likely to exceed the gravitational escape 
speed from the parent cloud (a few \kms ) and cannot be produced by 
infalling gas.  An alternative and more plausible scenario is that 
these three jet sources have been recently expelled from the cluster center by 
dynamical interactions such as the re-arrangement of non-hierarchal triple 
systems (cf. Sterzik \& Durisen 1995; Reipurth \etal\ 1999 and references therein).    
For a 10 \kms\ stellar motion through the medium, the time since ejection
from an interaction 0.3 parsec away (about the distance from the
cluster core) is about 30,000 years.  For an ensemble of 150 stars,
three such interactions over the 30,000 years is required to produce
the three C-shaped symmetric jets.

\section{Conclusions}

$\bullet$ 
We report the discovery of five irradiated jets in the
outskirts of the Orion Nebula (HH~502 through 506).  Four of 
these jets are at least partially photo-ionized by the Lyman continuum 
radiation field of the Trapezium stars.  All five jets exhibit C-shaped 
symmetry with jets and jet cocoons that bend away from the core of the 
Orion Nebula.  All five jets are intrinsically asymmetric with
one beam typically at least three times brighter than the counter
jet.  However, all jets do show evidence for being 
bipolar on larger scales.

$\bullet$
New observations of nine soft-UV irradiated jets are presented, including
five new discoveries (HH~495 -- 499), near the reflection nebula
NGC~1333 in Perseus.   These flows are irradiated by the non-ionizing UV
radiation field of B stars in the NGC~1333 cluster.  They differ from
most HH flows and jets in that their driving sources are visible
at visual wavelengths and that both beams are plainly seen with
relatively low obscuration.   As is the case for the irradiated jets
exposed to Lyman continuum radiation,  the circumstellar envelopes of
these young stars have been eroded, presumably by the soft UV radiation
field. 

$\bullet$ 
A model of soft-UV induced cloud core and protostellar 
envelope erosion is presented.   Apparently even non-ionizing radiation fields can 
expose the forming young stars and render their jets easily visible.
Dissociation,  heating, and photo-evaporation by soft UV may play 
a major role in destroying the envelopes from which young stars accrete, 
thereby bringing further stellar growth to a halt.  Irradiated jets 
are associated with naked young stars.   Their presence may signal one 
method by which Nature regulates the final masses of stars, namely
the termination of accretion by external photo-evaporation caused by
massive O, B, and even early A stars.  The photo-evaporative flows may
even play a fundamental role in the origins of turbulent motions inside 
molecular clouds.   

$\bullet$ 
A model for jet photo-ionization by Lyman continuum radiation is presented.  
Conical jets spreading with a finite opening angle become fully photo-ionized 
at some distance from the source.  We discuss several mechanisms which can 
explain the observed asymmetries of the 
irradiated jets.   Such asymmetries are likely to be produced by
dissimilar mass loading of the jet and counter jet which can produce
a kinematic asymmetry.   The resulting velocity difference between 
the jet and counter jet can make either the  fast beam brighter 
at a given distance from the source (if both beams spread at the 
{\it same} velocity orthogonal to the jet axis and thus have unequal 
opening angles) or can result in the slower beam being brighter (if 
the beams have the same opening angles).

$\bullet$ 
The physical mechanisms that can bend jets are discussed. In the Orion 
Nebula, C-shaped bending is likely to be a result of the rocket effect 
acting on the skin of the jet while it is only partially ionized. 
However, the jet cocoons may also be deflected by the ram pressure
of the champagne flow from the core of the nebula.  If magnetized, the
cocoons may also contribute to jet deflection.

$\bullet$ 
Three of the NGC~1333 jets exhibit C-shaped symmetry with bending
{\it towards} the cluster core.  We argue that these stars may
have been recently expelled from the cluster core by dynamical
interactions and are now moving through the surrounding medium at about
10 \kms . 

$\bullet$ 
Six new LL Orionis type wind-wind collision fronts are reported in the 
outskirts of the Orion Nebula.

\acknowledgments{
We thank David Devine for help with some of the observations and
data reduction.  We thank the referee,  Alex Raga, for a very
careful reading of the manuscript and for many valuable comments. 
We thank Taft Armandroff, Jim De Veny, and Frank Valdes for their
assistance with the MOSAIC and its software.  We thank the NOAO staff for
for their excellent support.  We acknowledge Matt Beasley for stressing  
the possible role of radiation pressure in bending irradiated jets.
This research was supported in part by NASA grant 
NAGW-3192 (LTSA) and in part by NSF grant NSF-9819820.}

\section{Appendix 1:  New Herbig-Haro Objects in Orion and NGC~1333}

In addition to the irradiated jets, our imaging survey has revealed 
additional Hebrbig-Haro objects which were identified as emission line 
features with compact structure.  Many, but not all, are most conspicuous 
in the [\sii ] images.  The HH numbers and positions of the objects are
listed in Table~6 and a short description of each object is given below. 

\noindent{\bf HH~547:} 
A very bright [\sii ] bow shock facing towards the
west and located about 10\arcmin\ west of the NGC~1333 cloud core.   A
faint envelope of emission can be traced back towards the east for about
an arc minute on the south side of the bow.

\noindent{\bf HH~542:}
A complex of [\sii ] knots and filaments that emerge from
a reflection nebula located about half an arc minute to the 
southeast of the HH objects.  The reflection nebula 
has an axis of symmetry at about PA = 315\arcdeg\ and 
points towards the HH~542 complex. 

\noindent{\bf HH~C1:}
A pair of bow shaped \Ha\ features located symmetrically on either side
of an \Ha\ emission line star.  The bows face away from this star.
A line drawn through the tips of the bows has PA = 10\arcdeg .
This is possibly an HH object.

\noindent{\bf HH~543:}
A small [\sii ] knot located just north of the HH~333 jet. There
is a hint of elongation along an east--west axis.

\noindent{\bf HH~544:}
A bright 5\arcsec\ long [\sii ] jet emerging towards PA = 100\arcdeg\ 
from a faint star embedded within the NGC~1333 reflection nebula.  
Two knots are resolved with the knot closest to the star being 
the brightest.

\noindent{\bf HH~C2:}
A chain of \Ha\ knots at PA = 10\arcdeg\ in the NGC~1333 reflection 
nebula.  The chain is located directly west of the bright star 
BD+30 549 and is embedded in its bright reflection nebulosity.
This is possibly an HH object.

\noindent{\bf HH~C3:}
A faint 10\arcsec\ long \Ha\ filament emerging from a star at PA = 
350\arcdeg . This feature may be a very faint irradiated jet.
There is a hint of a very faint counterjet.   There is an additional 
very faint \Ha\ filament located about 1\arcmin\ south of HH~C3
along the expected location of the couterjet. This feature also has 
an orientation of PA = 350\arcdeg .
This is possibly an HH object.

\noindent{\bf HH~496:}
A compact [\sii ] knot and a faint filament connecting it to
a star about 30\arcsec\ to the northwest.

\noindent{\bf HH~545:}
A large and diffuse [\sii ] bright bow shock facing east that is 
located about 7\arcmin\ east of BD+30 549.  The shock is about 
30\arcsec\ in diameter, has a sharp eastern edge, and a diffuse
western side.

\noindent{\bf HH~546:}
A large but faint  east-facing bow shock due east of HH~348 and 
349 (Bally \etal\ 1996a).  The shock can be traced for about 1\arcmin\
in the east-west direction and has a sharp northern edge.   This
structure appears to be a continuation of the outflow traced by   
HH348/349.   HH348, 349, and 546 may be powered by the deeply embedded 
infrared source IRAS 4B (Blake \etal\ 1995) which lies along the 
axis of symmetry of these HH objects. 

\noindent{\bf HH~384:} A 30\arcsec\ long filament at PA = 100\arcdeg\ in the
OMC2 region.  This object was discovered in the study of Yu \etal\ (1997) and
Reipurth \etal\ (1998).

\noindent{\bf HH~532:} A compact [\sii ] bright HH object located about
30\arcsec\ northwest of HH~506 several arc minutes north of M~43 and east of
OMC2.  The orientation and location of this bow shaped HH object suggests that 
it is unrelated to the nearby HH~506 irradiated jet.

\noindent{\bf HH~533:}  A 30\arcsec\ long [\sii ] filament oriented towards 
PA = 20\arcdeg\ located in the OMC2 region.  The morphology may be
indicative of a jet.

\noindent{\bf HH~535:}
A bright 30\arcsec\ long chain of compact bow shocks apparently propagating 
towards PA = 330\arcdeg\ from OMC2.  This collimated outflow appears to be aimed
at HH~44 located several arc minutes to the northwest,  and may represent 
either the driving jet, or a portion of the flow which excites that HH object.

\noindent{\bf HH~536:}
A cluster of faint [\sii ] knots and arcs about 1\arcmin\ northwest of
a compact reflection nebula whose axis of symmetry points towards HH~536. 
If this association is correct, then this flow lies parallel to but
slightly south and west of the HH~535/44 flow.

\noindent{\bf HH~537:} This is the brightest new HH object in the
survey fields, and one of the brightest HH objects in the sky.  Lying at 
the northern periphery of the M~42 \Hii\ region, this HH object has been 
missed because of the glare of the Orion Nebula.  This bright bow shock 
appears to be propagating towards PA = 235\arcdeg\ from the vicinity of 
OMC2 which is located about 7\arcmin\ towards the northeast (projected 
distance of about 1 pc).  HH~537 consists of a bright apex with a fainter 
and clumpy bow whose southern wing can be traced back for at least 1\arcmin .

\noindent{\bf HH~538:}  A bright but compact bow shock located a few arc minutes
north of M~43.  The morphology, which consists of a bright knot surrounded
by several fainter clumps in the wings of a bow shock suggests that this flow 
is propagating towards the southwest.

\noindent{\bf HH~539:}  A faint emission line knot located about 
1\arcmin\ northeast of HH~538.

\noindent{\bf HH~540:}  A pair of bright bow shocks located south of the
bright bar in the Orion Nebula.  The shocks are separated north-south
by about 40\arcsec .    The flow appears to be propagating from an unknown
source at PA = 200\arcdeg\ towards the south.

\noindent{\bf HH~541:}  A [\sii ] bright filament embedded within a 
complex filamentary nebula near the southern periphery of M~42 several
arc minutes south of the HH~504 irradiated jet.

\clearpage

\section*{References}

\noindent
Bally, J., \& Devine, D. 1994, ApJ, 428, L65  

\noindent
Bally, J., Devine, D., Fesen, R., \& Lane, A. P. 1995, ApJ, 454, 345

\noindent 
Bally, J., Devine, D., \& Reipurth, B.  1996a, ApJ, 473, 49. 

\noindent
Bally, J., Morse, J., \& Reipurth, B. 1996b
in {\it Science with Hubble Space Telescope - II}
eds. P. Benvenuti, F. D. Macchetto, \& E. J. Schreier, 
(STScI, Baltimore), p. 491

\noindent
Bally, J., O'Dell, C. R., \& McCaughrean, M. J. 2000, AJ (in press)

\noindent
Bally, J., \etal\  2000b, (in preparation).

\noindent
Blake, G. A.,  Sandell, G., van Dishoeck, E. F. 1995, ApJ, 441, 689 

\noindent
Devine, D., Bally, J., Reipurth, B., \& Heathcote, S.
1997, AJ, 114, 2095

\noindent
Cant\'o, J. \& Raga, A. C. 1995, MNRAS, 277, 1120

\noindent
C\u{e}rnis, K. 1990, Ap\&SS, 166, 315

\noindent
Chernin, L. \& Masson, C. R., Pino, E. G. D., \& Benz, W. 1994, ApJ, 426, 204

\noindent
Cernicharo, J., \etal\ 1998, Science, 282, 462 

\noindent
Gull, T. R., \& Sofia, S. 1979, ApJ, 230, 782

\noindent
Hartigan, P., Morse, J. A. \& Raymond, J.  1994, ApJ, 436, 125

\noindent
Herbig, G. H., \& Jones, B. F. 1983, AJ, 88, 1040

\noindent
Hillenbrand, L. A. 1997, AJ, 113, 1733

\noindent
Hillenbrand, L. A. \& Hartmann, L. W. 1998, ApJ, 492, 540

\noindent
Hirth, G. A., Mundt, R., Solf, J.  1997, A\&AS, 126, 437

\noindent 
Hollenbach, D. \&  Tielens, A. G. G. M. 1997, ARAA, 35, 179  

\noindent
Hollenbach, D. \&  Tielens, A. G. G. M. 1999, Rev.Mod.Phys. 71, 173 

\noindent 
Hollenbach, D., Johnstone,, D., Lizano, S. \& Shu, F. H. 1994, ApJ, 428, 654 

\noindent
Johnstone,, D.,  Hollenbach, D.,  \& Bally, J. 1998, ApJ, 499, 758

\noindent
Lada, C. J., Alves, J., \& Lada, E. A. 1996, AJ, 111, 1964

\noindent
Masson, C. R., \& Chernin, L. M. 1993, ApJ, 414, 230

\noindent
McCaughrean, M. J. \& Stauffer, J. R. 1994, AJ, 108, 1382

\noindent
O'Dell, C. R. 1994, \apss, 216, 267 

\noindent
O'Dell, C. R., Valk, J. H., Wen, Z.  \& Meyer, D. M. 1993, \apj, 403, 678 

\noindent
Wen, Z.  \& O'Dell, C. R. 1993, \apj, 409, 262 

\noindent
Raga, A., L\'opez-Mart\'in, L., Binette, L., Lo\'pez, J. A., 
 Cant\'o, J., Arthur, A. J., Mellema, G., 

Steffen, W., \& Ferruit, P.  2000, MNRAS, 314, 681. 

\noindent
Raga, A, \& Cant\'o, J. 1989, PASP, 111, 1964

\noindent
Raga, A., \& Cabrit, S. 1993, A\&A, 278, 267

\noindent
Reipurth, B. 1999, {\it A General Catalogue of Herbig-Haro Objects},
 2. Edition, 
published electronically at http://casa.colorado.edu/hhcat

\noindent
Reipurth, B., Bally, J., \& Devine, D. 1997, AJ, 114, 2708

\noindent
Reipurth, B., Bally, J., Fesen, R., \& Devine, D. 1998, Nature, 396, 343

\noindent
Reipurth, B., Yu, K. C., Rodr\'iguez, L. F., Heathcote, S., \& Bally, J.
1999, A\&A,  352, L83 

\noindent
Sterzik, M. F., \& Durisen, R. H., 1995, AA, 304, L9

\noindent
Yu, K. C., Bally, J., \& Devine, D.
1997, ApJL, 485, 45


\newpage
\clearpage
\small
\refstepcounter{table}
\label{Table 1}
\begin{tabular}[]{ l l c c c }

\multicolumn{3}{l}{\sc Table 1: Journal of Spectroscopic Observations} \\
\hline
Object &  HH    & Date  & PA        & Comments \\
       &        &       & (\arcdeg ) &  \\
\hline
\hline
NGC 1333 &  333     &  1/11/96 &  74 & 3 $\times$ 500s  \\         
NGC 1333 &  334     & 12/27/98 &115.7& 3 $\times$ 600s  \\ 
NGC 1333 &  336     &  1/11/96 & 135 & 3 $\times$ 1000s  \\   
NGC 1333 &  499 s1  & 12/27/98 &  25 & 3 $\times$ 600s star+jet \\ 
NGC 1333 &  499 s2  & 12/27/98 &  25 & 3 $\times$ 600s knots NE1 and S1 \\
M42      &  502 s1  & 12/27/98 &  22 & 2 $\times$ 100s   \\  
M42      &  502 s2  & 12/27/98 &  40 & 2 $\times$ 100s   \\  
M42      &  502 s3  & 12/27/98 &   7 & 3 $\times$ 600s   \\  
M42      &  503 s1  & 12/29/98 & 113 & 3 $\times$ 900s star + jet  \\ 
M42      &  503 s2  & 12/29/98 & 113 & 3 $\times$ 900s 9.8\arcsec\ S of s1 \\ 
M42      &  503 s3  & 12/29/98 & 120 & 3 $\times$ 900s 26.4\arcsec\ S of s1. \\ 
\hline
\end{tabular}

\footnotesize
\smallskip
\bigskip
\clearpage

\clearpage
\small
\refstepcounter{table}
\begin{tabular}[]{ l c c c l }

\multicolumn{5}{l}{\sc Table 2a: Irradiated Jets in M42 } \\
\hline
Feature    & Location           & PA & $\theta _{def}$    & Comments \\

     & $\alpha (J2000)$  $\delta (J2000)$  & (\arcdeg ) & (\arcdeg ) & \\
\hline
\hline
HH502 S8     & 5 35 25.1,  -5 31 39    & 196 & 12 &  \\
HH502 S7     & 5 35 26.0,  -5 30 54    & 200 &  8 &  \\
HH502 S6     & 5 35 26.1,  -5 30 39    & 202 &  6 &  \\
HH502 S5     & 5 35 26.6,  -5 30 17    & 203 &  5 &  \\
HH502 S4     & 5 35 27.0,  -5 30 04    & 204 &  4 &  \\
HH502 S3     & 5 35 27.1,  -5 29 59    & 206 &  2 &  \\
HH502 S2     & 5 35 27.2,  -5 29 53    & 206 &  2 &  \\
HH502 S1     & 5 35 27.5,  -5 29 45    & 207 &  1 &  \\
HH502 jet    & 5 35 27.9,  -5 29 35    & 208 & -- \\ 
HH502 star   & 5 35 28.0,  -5 29 32    &     & -- \\ 
HH502 N1     & 5 35 28.1,  -5 29 28    & 27  & -- &  \\
HH502 N2     & 5 35 30.1,  -5 28 55    & 39  & 12 &  \\
HH502 N3     & 5 35 30.8,  -5 28 43    & 41  & 14 &  \\
HH502 N3 wing& 5 35 30.7,  -5 28 50    & -   & -- \\
   & & & & \\
HH~503 W2    & 5 34 43.6,  -5 31 28    & 283 & 9 &   \\ 
HH~503 W1    & 5 34 44.8,  -5 31 32    & 284 & 8 &    \\ 
HH~503 jet   & 5 34 48.4,  -5 31 43    & 296 & - &   \\ 
HH~503 star  & 5 34 48.9,  -5 31 46    &     & & C \\ 
HH~503 E1    & 5 34 51.6,  -5 32 13    & 123 & 8 &   \\
HH~503 E2w   & 5 34 54.8,  -5 33 06    & 133 &18&   \\
HH~503 E2e   & 5 34 55.0,  -5 33 07    & 133 &18&   \\
   & & & & \\
HH~504 star  & 5 35 03.6,  -5 29 27    &     &   & C \\ 
HH~504 jet   & 5 35 03.5,  -5 29 22    & 340 & 0 &   \\ 
HH~504 N1    & 5 35 03.0,  -5 29 00    & 340 & 0 &   \\ 
HH~504 N2    & 5 35 02.8,  -5 28 55    & 337 & 3 &   \\ 
HH~504 bow   & 5 35 02.2,  -5 28 47    & 330 & 10&   \\ 
   & & & & \\
HH~505 star  & 5 34 40.9,  -5 22 43    &     & & C \\ 
HH~505 jet   & 5 34 40.8,  -5 22 39    & 342 &  0 &   \\ 
HH~505 N1    & 5 34 40.8,  -5 22 35    & 341 &  1 &   \\ 
HH~505 N2    & 5 34 39.6,  -5 22 07    & 330 & 12 &   \\ 
HH~505 N3    & 5 34 39.4,  -5 22 04    & 330 & 12 &   \\ 
   & & & & \\
HH~506 S6    & 5 35 47.5,  -5 10 54    & 175 &15 &   \\ 
HH~506 S5    & 5 35 47.5,  -5 10 51    & 177 &13 &   \\ 
HH~506 S4    & 5 35 47.4,  -5 10 48    &   3 & 7 &   \\ 
HH~506 S3    & 5 35 47.3,  -5 10 44    &   9 & 1 &   \\ 
HH~506 S2    & 5 35 47.3,  -5 10 40    &  10 & 0 &   \\ 
HH~506 S1    & 5 35 47.4,  -5 10 37    &  10 & 0 &   \\ 
\hline
\end{tabular}
\clearpage
\small
\refstepcounter{table}
\label{Table 2a}
\begin{tabular}[]{ l c c c l }

\multicolumn{5}{l}{\sc Table 2a Continued} \\
\hline
Feature    & Location           & PA & $\theta _{def}$    & Comments \\

     & $\alpha (J2000)$  $\delta (J2000)$  & (\arcdeg ) & (\arcdeg ) & \\
\hline
\hline
HH~506 star  &  5 35 47.4,  -5 10 29  &  10 & 0 &   \\ 
HH~506 N1    &  5 35 47.6,  -5 10 25  &  10 & 0 &   \\
HH~506 N2    &  5 35 47.8,  -5 10 21  &  10 & 0 &   \\
HH~506 N3    &  5 35 48.6,  -5 10 16  &  10 & 0 &   \\
HH~506 N4    &  5 35 49.0,  -5 10 07  &  10 & 0 &   \\
\hline

\end{tabular}

\footnotesize
\smallskip
\noindent Notes:
1].  C refers to C-shaped symmetry.
     S refers to S-shaped symmetry.
2]  $\theta _{def}$ is the difference between the position angle of 
the jet near the source and the position angle of a line connecting 
the listed feature to the source.


\small
\refstepcounter{table}
\begin{tabular}[]{ l c c c l }

\multicolumn{5}{l}{\sc Table 2b: Irradiated Jets in NGC~1333} \\
\hline
Feature    & Location           & PA & $\theta _{def}$    & Comments \\

     & $\alpha (J2000)$  $\delta (J2000)$  & (\arcdeg ) & (\arcdeg ) & \\
\hline
\hline
HH~333 W6    & 3 28 52.6,  31 25 56 & 268& 15& diffuse  \\
HH~333 W5    & 3 28 57.3,  31 25 35 & 260& 7 & diffuse  \\
HH~333 W4    & 3 29 02.4,  31 25 46 & 259& 6 & chain    \\
HH~333 W3    & 3 29 03.6,  31 25 46 & 258& 5 & compact  \\
HH~333 W2    & 3 29 06.2,  31 25 50 & 254& 1 & compact  \\
HH~333 W1    & 3 29 10.9,  31 26 03 & 253& 0 & compact  \\
HH~333 star  & 3 29 11.7,  31 26 10 & 73 & --&  S \\ 
HH~333 E1    & 3 29 13.4,  31 26 15 & 73 & 0 & compact   \\
HH~333 E2    & 3 29 15.6,  31 26 25 & 73 & 0 & compact   \\
HH~333 E3    & 3 29 18.5,  31 26 37 & 73 & 0 & small bow \\

 & & & & \\

HH~334 W1    & 3 28 56.7,  31 22 01 & 292 & 3  & double knot  \\ 
HH~334 star  & 3 28 59.5,  31 21 47 & 115 & -- &  C \\ 
HH~334 E1    & 3 29 02.1,  31 21 28 & 120 & 5  & compact   \\ 
HH~334 E2    & 3 29 08.6,  31 20 34 & 120 & 5  & double knot    \\ 

 & & & & \\

HH~335 N3    & 3 29 11.9,  31 24 33 & 327& 21 & very faint  \\
HH~335 N2    & 3 29 15.9,  31 23 47 & 342& 6  & faint knot  \\
HH~335 N1    & 3 29 16.7,  31 23 30 & 345& 3  & bright knot \\
HH~335 jet   & 3 29 17.6,  31 22 59 & 348& -- & jet \\
HH~335 star  & 3 29 17.8,  31 22 46 & -- & -- & C(?) Bent jet in refl. neb. \\

 & & & & \\

HH~336 NW5   & 3 29 35.6,  31 19 25 & 335 & 21 & bow \\ 
HH~336 NW4   & 3 29 35.2,  31 19 02 & 326 & 12 &  \\ 
HH~336 NW3   & 3 29 35.3,  31 18 45 & 320 & 6  &  \\ 
HH~336 NW2   & 3 29 36.8,  31 18 20 & 316 & 2  & bow \\ 
HH~336 NW1   & 3 29 38.5,  31 17 57 & 314 & 0  & diffuse \\ 
HH~336 star  & 3 29 39.6,  31 17 44 & 314 & -- & S \\ 
HH~336 SE1   & 3 29 41.0,  31 17 29 & 134 & 0 &  \\ 
HH~336 SE2   & 3 29 46.6,  31 16 10 & 138 & 4 & faint [\sii ] \\ 

 & & & & \\

HH~495 star  & 3 28 46.3,  31 16 38 & --  & -- & one-sided  \\
HH~495 S1    & 3 28 46.0,  31 16 28 & 10  & 0  & one-sided  \\
HH~495 S2    & 3 28 45.9,  31 16 24 & 10  & 0  & one-sided  \\
HH~495 S3    & 3 28 45.7,  31 16 11 & 10  & 0  & one-sided  \\
HH~495 S4    & 3 28 45.6,  31 16 07 & 10  & 0  & one-sided  \\

 & & & & \\

HH~497 S3    & 3 29 51.7,  31 19 49 & 195& 5 &  \\ 
HH~497 S2    & 3 29 52.1,  31 19 51 & 193& 3 &  \\ 
HH~497 S1    & 3 29 53.0,  31 20 21 & 190& 0 &  \\ 
HH~497 star  & 3 29 54.0,  31 20 54 & -- & -- & S \\ 
HH~497 N1    & 3 29 54.5,  31 21 13 & 13 & 0 &  \\ 
HH~497 N2    & 3 29 55.2,  31 21 47 & 13 & 0 &  \\
HH~497 N3    & 3 29 55.9,  31 21 54 & 20 & 7 &  \\

\hline
\end{tabular}

\clearpage
\small
\refstepcounter{table}
\label{Table 2b}
\begin{tabular}[]{ l c c c l }

\multicolumn{5}{l}{\sc Table 2b Continued} \\
\hline
Feature    & Location           & PA & $\theta _{def}$    & Comments \\

     & $\alpha (J2000)$  $\delta (J2000)$  & (\arcdeg ) & (\arcdeg ) & \\
\hline
\hline
HH~498 SW1   & 3 28 51.4,  31 22 34 & 220 & 40 &   \\
HH~498 jet   & 3 28 51.7,  31 22 44 & 260 & -- & C \\ 
HH~498 star  & 3 28 52.2,  31 22 45 & --  & -- & C \\ 
HH~498 SE1   & 3 28 53.7,  31 22 36 & 115 & 35 &   \\

 & & & & \\

HH~499 S1    & 3 29 02.9,  31 22 12 & 185& 24 &   \\
HH~499 S jet & 3 29 02.9,  31 22 34 & 209& -- & C \\ 
HH~499 star  & 3 29 03.1,  31 22 38 & -- & -- & C \\ 
HH~499 N jet & 3 29 03.3,  31 22 43 & 29 & -- & C \\ 
HH~499 NE1   & 3 29 04.7,  31 23 01 & 43 & 14 &   \\
HH~499 NE2   & 3 29 06.4,  31 23 12 & 50 & 21 &   \\

\hline
\end{tabular}

\footnotesize
\smallskip
\noindent Notes:
1].  C refers to C-shaped symmetry.
     S refers to S-shaped symmetry.
2] The angle $\theta _{def}$ is the deflection angle.  It is
the absolute value of the difference between the position angle 
of the jet near the star and the position angle of a line drawn from
the star through the feature in question.    


\clearpage
\small
\refstepcounter{table}
\label{Table 3}
\begin{tabular}[]{ l c c c c c c}

\multicolumn{7}{l}{\sc Table 3: Outflow Properties} \\
\hline
Feature    & I(\Ha )$^1$ & EM & L & $n_e$ & I([\sii ])$^1$ & I(\Ha )/I([\sii ])  \\
  & ($\times 10^{-15}$)  & ($\rm cm^{-6} pc$) & ($\times 10^{-3}$pc) & 
  ($\rm cm^{-3}$) & ($\times 10^{-16}$) &    \\
\hline
\hline
HH502 N3 wing & 12.3  & 6055  & 4.48  & 1163  &   8.3  & 14.8  \\
~~~~~N3      & 17.5  & 8570  & 2.24  & 1956  &  15.5  &  11.3 \\
~~~~~N2      &  8.6  & 4192  & 2.24  & 1368  &   6.0  &  14.3 \\
~~~~~N1      &  0.95 &  466  & 2.24  &  456  &$<$2.4  &$>$4.0 \\
~~~~~jet     &  8.2  & 4005  & 1.12  & 1891  &   3.6  &  23.0 \\
~~~~~S1      &  3.8  & 1863  & 1.12  & 1290  &$<$2.4  &$>$15.0\\
~~~~~S2      &  3.4  & 1640  & 1.78  &  960  &$<$2.4  &$>$13.0\\
~~~~~S3      &  3.8  & 1863  & 2.24  &  912  &   3.6  &  10.6 \\
~~~~~jet segment  &  0.95 &  466  & 1.78  &  512  &$<$2.4  &$>$3.0 \\
~~~~~S5      &  9.5  & 4658  & 2.24  & 1442  &  12.0  &   8.0 \\ 
~~~~~S6      & 13.1  & 6428  & 2.24  & 1694  &  31.1  &   4.2 \\
~~~~~S7      &  5.5  & 2702  & 3.35  &  898  &   2.4  &  23.0 \\
~~~~~s8      &  0.95 &  466  & 4.47  &  323  &$<$1.2  &$>$10.0\\
HH503 W2     &  2.2  & 1053  & 2.24  &  686  &   8.4  &   2.6 \\
~~~~~W1      &  1.96 &  959  & 1.57  &  782  &   4.8  &   4.1 \\
~~~~~jet     &  0.95 &  466  & 1.12  &  645  &   4.8  &   2.1 \\
~~~~~E1      &  3.24 & 1584  & 1.12  & 1189  &   2.4  &  13.0 \\
~~~~~E2w     &  3.24 & 1584  & 1.12  & 1189  &  15.5  &   2.1 \\
~~~~~E2e     &  4.19 & 2049  & 1.12  & 1353  &  24.0  &   1.7 \\
HH504 jet    &  1.33 &  652  & 1.57  &  644  &   4.8  &   2.8 \\
~~~~~bow     &  3.81 & 1863  & 1.57  & 1089  &  14.4  &   2.7 \\
HH505 jet    &  3.81 & 1863  & 1.12  & 1290  &   8.3  &   4.7 \\
~~~~~N1      &  9.52 & 4658  & 1.12  & 2039  &  41.9  &   2.3 \\
~~~~~bow     & 15.24 & 7452  & 8.96  &  912  &  26.3  &   5.8 \\
~~~~~N2      & 17.15 & 8384  & 3.36  & 1580  &  17.9  &   9.5 \\
~~~~~N3      & 15.81 & 7732  & 1.57  & 2219  &  41.9  &   3.8 \\
HH506 S1     &$<$0.95&$<$466  & 1.57 &$<$545  &   6.0  & $<$1.3\\
~~~~~S2      &   -   &   -   &   -   &   -   &   2.4  &   -   \\
~~~~~S3      &   -   &   -   &   -   &   -   &   3.6  &   -   \\
~~~~~S4      &   -   &   -   &   -   &   -   &   4.8  &   -   \\
~~~~~S5      &   -   &   -   &   -   &   -   &   2.4  &   -   \\
~~~~~S6      &   -   &   -   &   -   &   -   &   3.6  &   -   \\ 
~~~~~N1(jet) &   -   &   -   &   -   &   -   &   7.2  &   -   \\ 
~~~~~N2      &   -   &   -   &   -   &   -   &   1.2  &   -   \\ 
~~~~~N3      &   -   &   -   &   -   &   -   &   1.2  &   -   \\ 
~~~~~N4      &   -   &   -   &   -   &   -   &   1.2  &   -   \\
\hline
\end{tabular}

\footnotesize \smallskip
\noindent Notes:
(1). Units are $\rm erg s^{-1} cm^{-2} arcsec ^{-2}$.
(2). The calibration has been bootstrapped from HST images of the
Orion Nebula.  The surface brightness is computed from the KPNO 4 meter
reduced images from the following relationships:
I(\Ha ) = $\rm 1.143 \times 10^{-14} * COUNTS / EXPTIME$
   $\rm (erg s^{-1} cm^{-2} arcsec^{-2})$. 
The emission measure then is given by 
$\rm EM(\Ha ) = 4.89 \times 10^{17} * I(\Ha )$
   $\rm (cm^{-6} pc)$.
Therefore, the electron density is given by
$\rm n_e = [ EM / L(pc) ]^{0.5}$
   $\rm (cm^{-3})$. 
  
\clearpage

\newpage
\clearpage
\small
\refstepcounter{table}
\label{Table 4}
\begin{tabular}[]{ l c c c c c c }

\multicolumn{7}{l}{\sc Table 4: Mechanical Parameters of the Ionized
Jets} \\
\hline
Flow & $r_j$ & $v_j$ & $n_j$ & $\dot M$ &  $\dot P$  & $\dot E$ \\

     & (AU)  &(\kms ) &($cm^{-3}$)& ($M_{\odot} ~yr^{-1}$) 
             & {\rm ($M_{\odot} yr^{-1} km s^{-1}$)} & ($L_{\odot}~yr^{-1}$)  \\
\hline
\hline
HH502 & 115 & 100 & 1891 & $\rm 6.5\times 10^{-9}$ & $\rm 6.5 \times 10^{-7}$         & $\rm 5.1 \times 10^{-3}$  \\
HH503 & 115 & 100 &  645 & $\rm 2.2\times 10^{-9}$ & $\rm 2.2 \times 10^{-7}$         & $\rm 2.2 \times 10^{-3}$  \\
HH504 & 115 & 100 & 1891 & $\rm 2.2\times 10^{-9}$ & $\rm 2.2 \times 10^{-7}$         & $\rm 1.7 \times 10^{-3}$  \\
HH505 & 115 & 100 & 1891 & $\rm 4.4\times 10^{-9}$ & $\rm 4.4 \times 10^{-7}$         & $\rm 3.4 \times 10^{-3}$  \\
 \hline

\end{tabular}

\footnotesize
\smallskip
\noindent Notes:
1].  The jet radius is an upper bound since the beams are not
resolved.   The velocity is assumed to be 100 \kms .
The computation of the mass loss parameters assumes that the
jet beam is fully ionized.  If it is not, these parameters 
only apply to the ionized parts, and are then lower bounds.
    

\newpage
\small
\refstepcounter{table}
\label{Table 5}
\begin{tabular}[]{ l c c c }

\multicolumn{3}{l}{\sc Table 5: LL Orionis Type Bow Shocks in M42 } \\
\hline
Feature    & Location                          & PA        & Comments \\

	   & $\alpha (J2000)$  $\delta (J2000)$  & (\arcdeg) & \\
\hline
\hline
 LL1    &  5 35 05.6,  -5 25 20  & 75  & LL Ori, W of M42 core \\ 
 LL2    &  5 34 40.8,  -5 22 43  & 90  & HH~505                \\ 
 LL3    &  5 34 40.8,  -5 26 39  & 75  &                       \\ 
 LL4    &  5 34 42.7,  -5 28 38  & 45  &                       \\ 
 LL5    &  5 35 31.5,  -5 28 16  & 340 & in HH~400             \\ 
 LL6    &  5 35 32.9,  -5 30 21  & 5   & in HH~400             \\ 
 LL7    &  5 35 35.1,  -5 33 49  & 345 & S of Bar              \\ 
\hline

\end{tabular}

\footnotesize
\smallskip
\noindent 

\normalsize

\newpage
\small
\refstepcounter{table}
\label{Table 6}
\begin{tabular}[]{ l c l }

\multicolumn{3}{l}{\sc Table 6: New Herbig-Haro Objects in NGC~1333 and M42} \\
\hline
Name      &       Location                      & Comments \\
	  & $\alpha (J2000)$  $\delta (J2000)$  &          \\
\hline
\hline
 HH~547   &  3 27 54.5, 31 09 36.9  & Very bright [Sii] bow facing west \\
 HH~542   &  3 28 49.5, 31 18 47.5  & Amorphous flow at PA = 315\arcdeg\ \\ 
 HH~C1    &  3 29 04.3, 31 19 06.4  & Bipolar \Ha\ flow from star at PA = 10\arcdeg\ \\
 HH~543   &  3 29 14.7, 31 27 21.9  & Knot north of HH~333         \\ 
 HH~544   &  3 29 14.3, 31 22 47.4  & Compact [\sii] jet in N~1333 \\
 HH~C2    &  3 29 18.5, 31 24 46.8  & Chain of \Ha\ knots at PA=10 near BD+30 549 \\
 HH~C3    &  3 29 30.5, 31 19 04.1  & Candidate \Ha\ jet at PA = 350\arcdeg\ from star \\
          &  3 29 33.4, 31 16 59.2  & \Ha\ filament (counter jet to C3 ?) \\ 
 HH~496   &  3 29 34.7, 31 24 22.3  & [\sii ] knot east of BD+30 549 \\
 HH~545   &  3 29 52.4, 31 24 31.4  & Large east facing [\sii ] bow \\
 HH~546   &  3 29 58.4, 31 13 59.9  & Large east facing [\sii] bow  \\ 
	  &                         &  \\
 HH~537   &  5 35 01.3, -5 14 07.1  & Very bright bow, N rim of M~42 \\ 
 HH~541   &  5 35 06.0, -5 33 30.2  & Bright knots south of HH~504 \\ 
 HH~536   &  5 35 18.4, -5 12 45.6  & Faint [\sii ] knots, arcs     \\ 
 HH~540A  &  5 35 19.5, -5 31 06.3  & Bow shock west of HH~502     \\ 
 HH~540B  &  5 35 18.4, -5 31 42.7  & Second bow shock west of HH~502 \\ 
 HH~535   &  5 35 19.0, -5 11 40.4  & Jet driving HH~44?           \\ 
 HH~384   &  5 35 25.9, -5 09 23.1  & E-W chain of [\sii ] knots   \\ 
 HH~533   &  5 35 29.5, -5 10 01.7  & Faint [\sii ] jet?            \\ 
 HH~538   &  5 35 33.3, -5 13 09.5  & Bow N of M~43                \\ 
 HH~539   &  5 35 37.0, -5 11 43.8  & Faint knot NE of HH~538      \\ 
 HH~532   &  5 35 46.1, -5 09 53.6  & NW of HH~506                 \\ 
\hline

\end{tabular}

\footnotesize
\smallskip
\noindent 

\normalsize


\begin{figure*}[tbp]
\includegraphics[height=7.5in]{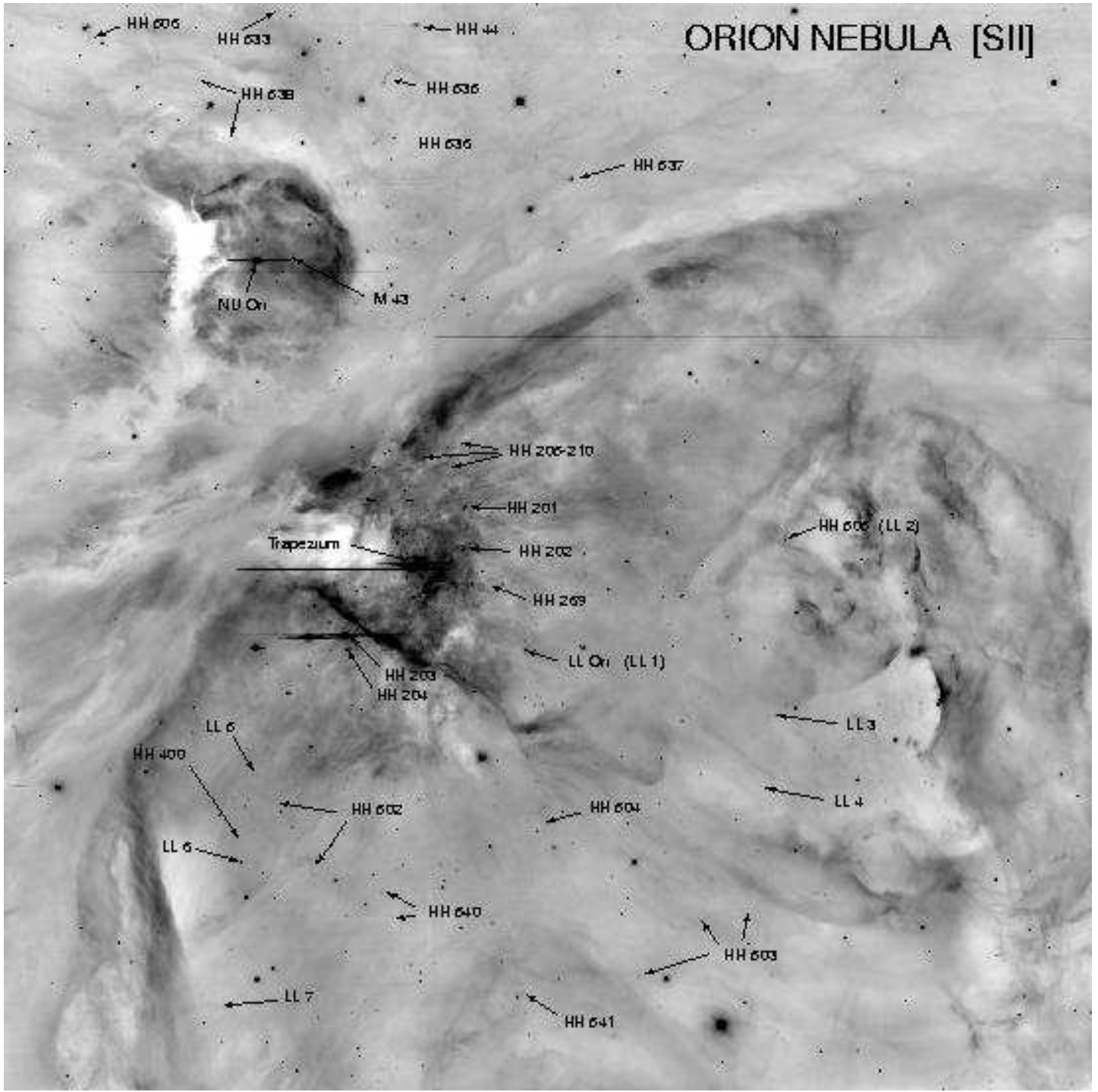}
\caption[]{
A sub-field of the MOSAIC 4 meter prime focus [\sii ] image of the 
Orion Nebula from which the large scale intensity gradients have been 
removed.  The locations of the various jets, HH objects,  and LL Ori 
type wind-wind collision fronts are marked. 
The image is an 80\% unsharp mask of the logarithm of 
the  pixel values (`soft unsharp mask').  To make the mask, 
the logarithmic image was clipped to eliminate bright stars
and a 2 dimensional 7-th order polynomial function was fit to the image
intensity surface.  After scaling the mask by an amplitude factor of 
0.80, it was subtracted from the logarithmic image. North is at the 
top and east is to the left. The image is 26.0\arcmin\ in extent in 
the horizontal (E--W) direction and 28.2\arcmin\ in extent in the 
vertical direction.
}
\end{figure*}

\begin{figure*}[tbp]
\includegraphics[height=6.5in]{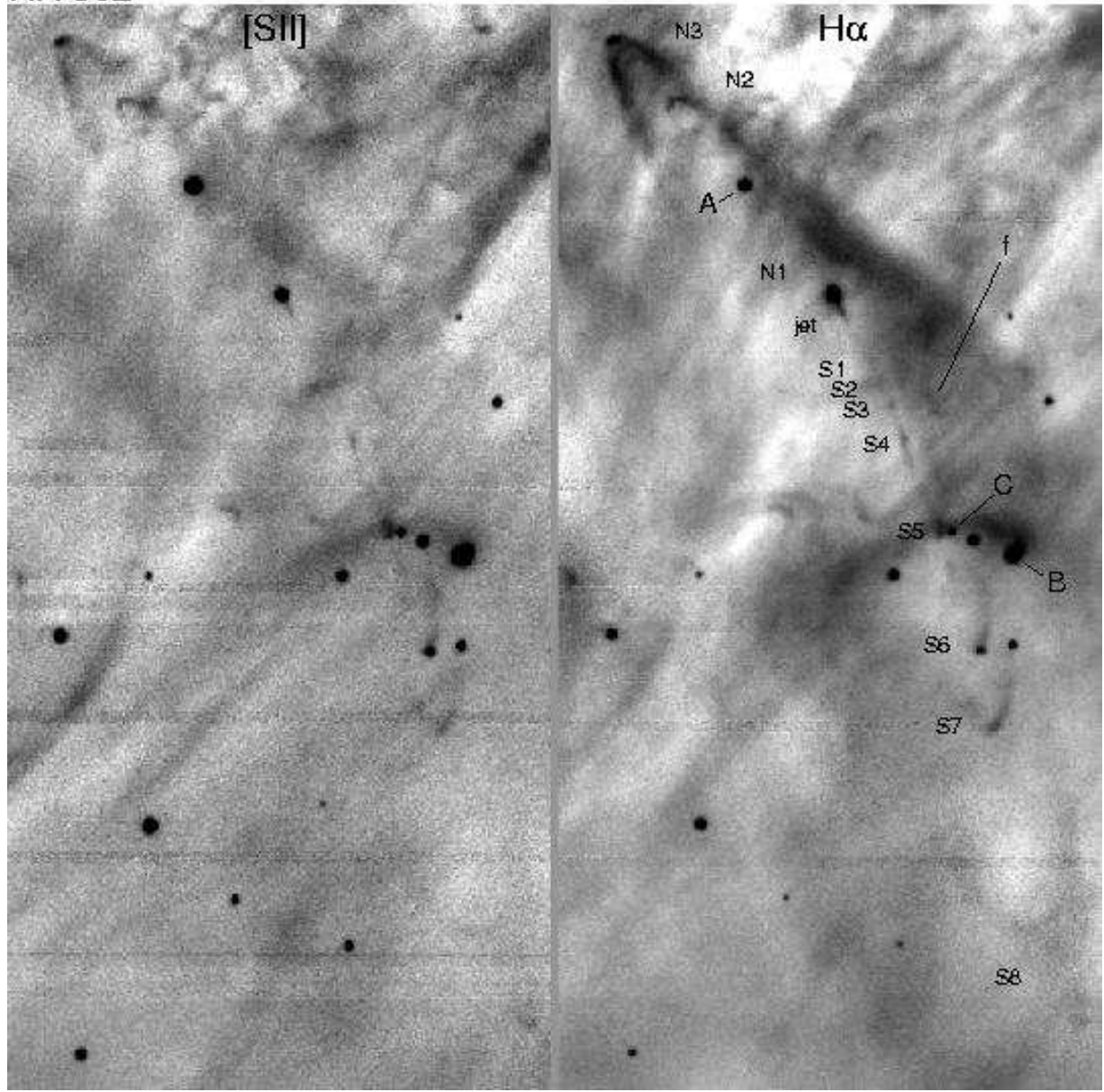}
\caption[]{
HH 502 in \Ha\ (right) and [\sii ] (left).  
The intensity gradient produced by the Orion Nebula has been removed by
subtracting a third order polynomial function fit to the background data.
North is towards the top and east is towards the left in all
illustrations.
} 
\end{figure*}

\begin{figure*}[tbp]
\includegraphics[height=6.5in]{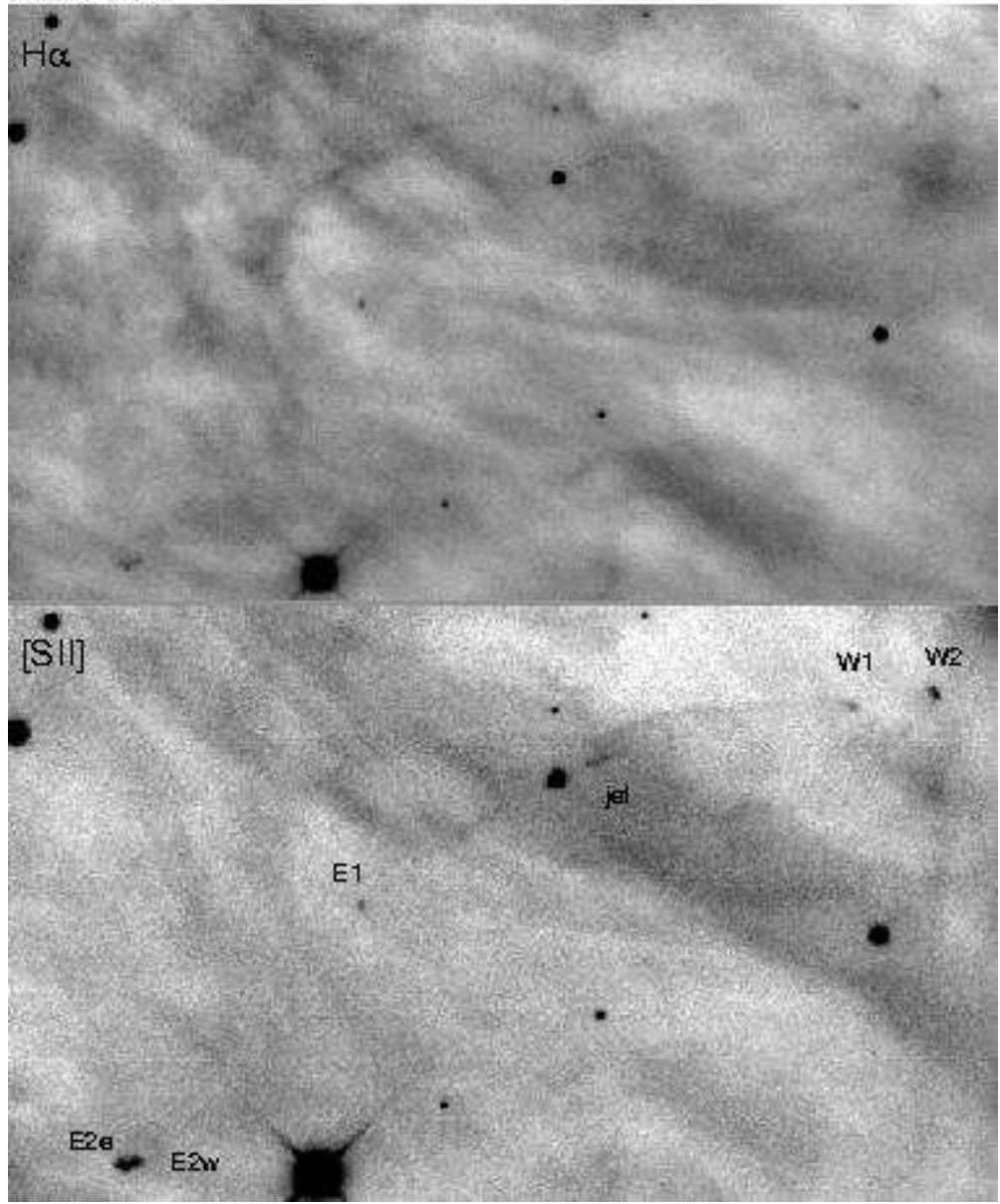}
\caption[]{
HH 503 in the southern portion of the Orion Nebula in 
\Ha\ (top) and [\sii ] (bottom).  
} 
\end{figure*}

\begin{figure*}[tbp]
\includegraphics[height=4.0in]{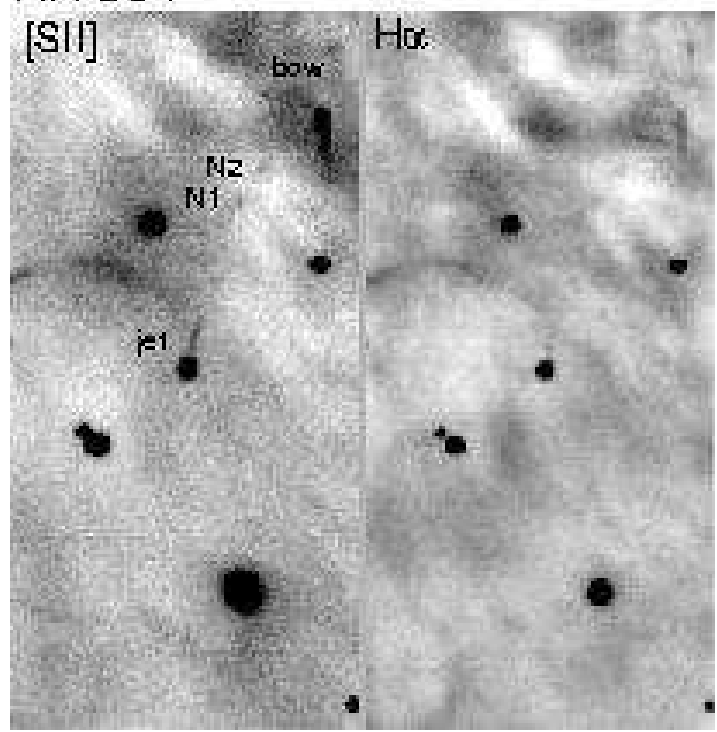}
\caption[]{
HH 504 with \Ha\ (right) and [\sii ] (left). 
} 
\end{figure*}

\begin{figure*}[tbp]
\includegraphics[height=4.0in]{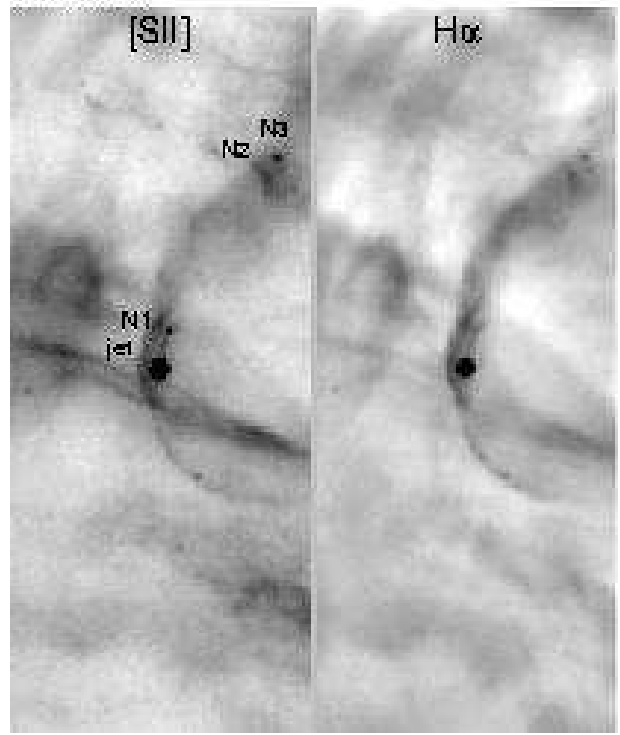}
\caption[]{
HH 505 with \Ha\ (right) and [\sii ] (left). 
} 
\end{figure*}

\begin{figure*}[tbp]
\includegraphics[height=4.0in]{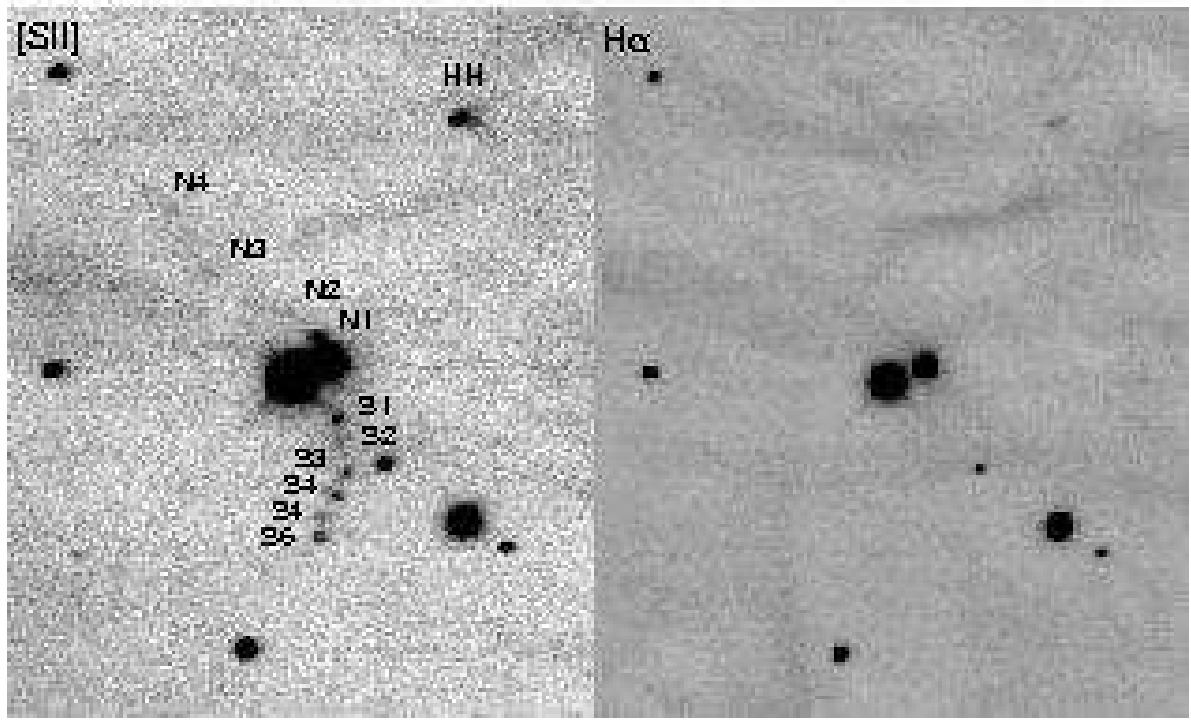}
\caption[]{
HH 506 with \Ha\ (right) and [\sii ] (left). 
} 
\end{figure*}

\begin{figure*}[tbp]
\includegraphics[height=7.5in]{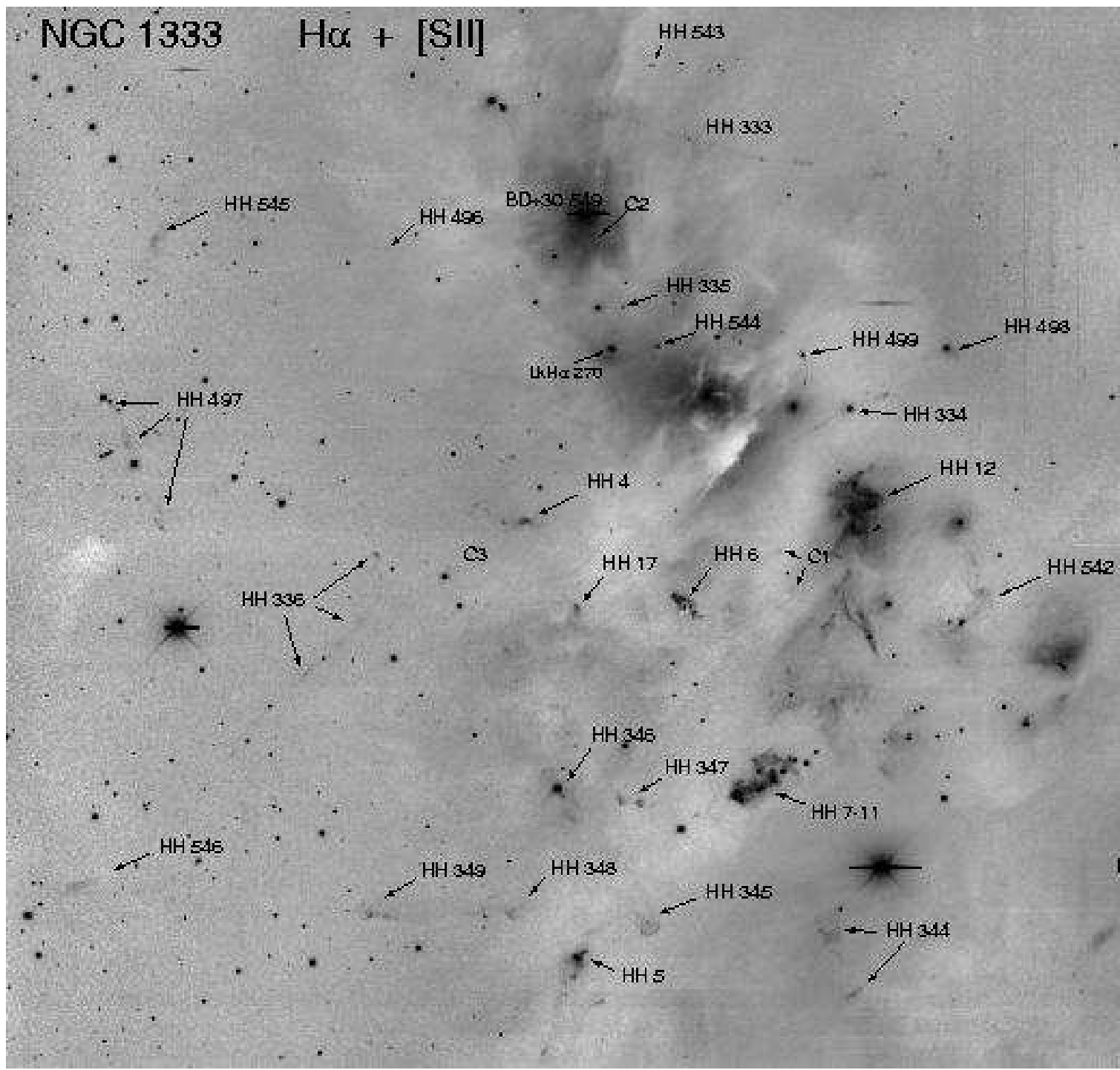}
\caption[]{
An \Ha\ + [\sii ] composite image of the NGC~1333 region
obtained with the MOSAIC camera at the prime focus of
the Mayall 4 meter telescope.   The locations 
of the various jets and other HH objects in the field are
indicated.   The image was processed in the same manner as
Figure~1.  North is at the top and east is to the left. 
The image is 19.5\arcmin\ in extent in the horizontal direction
and 17.3\arcmin\ in extent in the vertical direction.
} 
\end{figure*}

\begin{figure*}[tbp]
\includegraphics[height=7.0in,angle=-90]{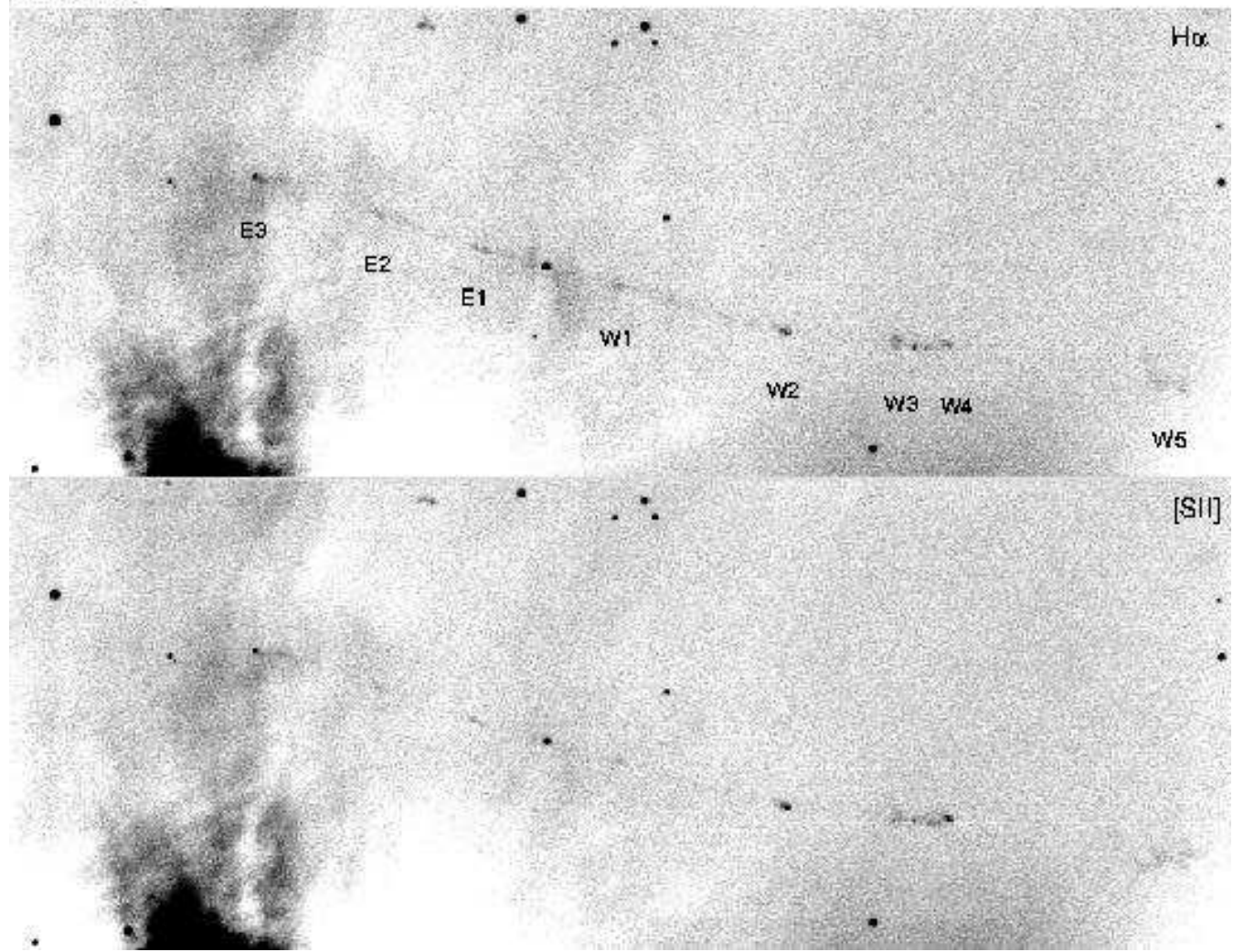}
\caption[]{
HH~333 in NGC~1333 in \Ha\ (top) and [\sii ] (bottom).
} 
\end{figure*}

\begin{figure*}[tbp]
\includegraphics[height=6.5in]{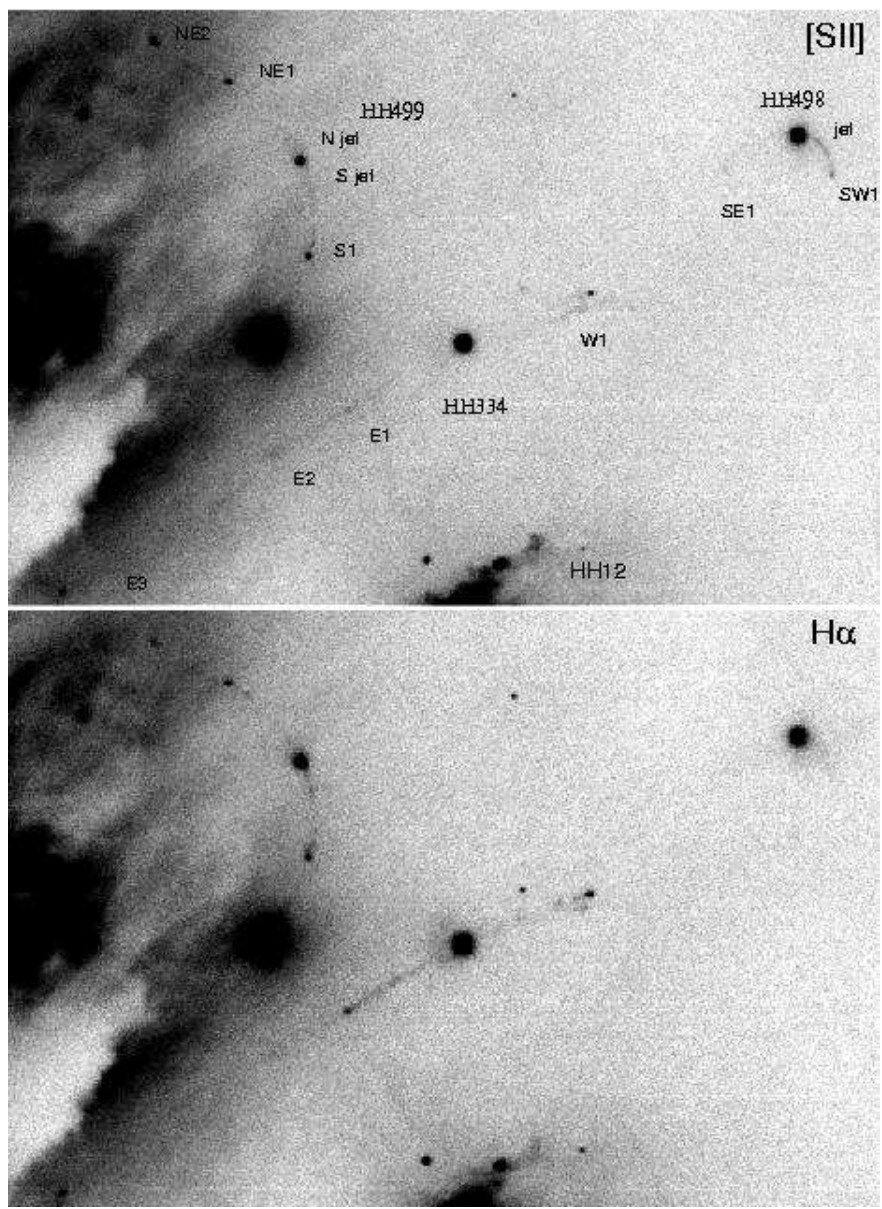}
\caption[]{
HH~334, HH~498, and  HH~499 in NGC~1333 in [\sii ] (top) and \Ha\ (bottom).
} 
\end{figure*}

\begin{figure*}[tbp]
\includegraphics[height=6.5in]{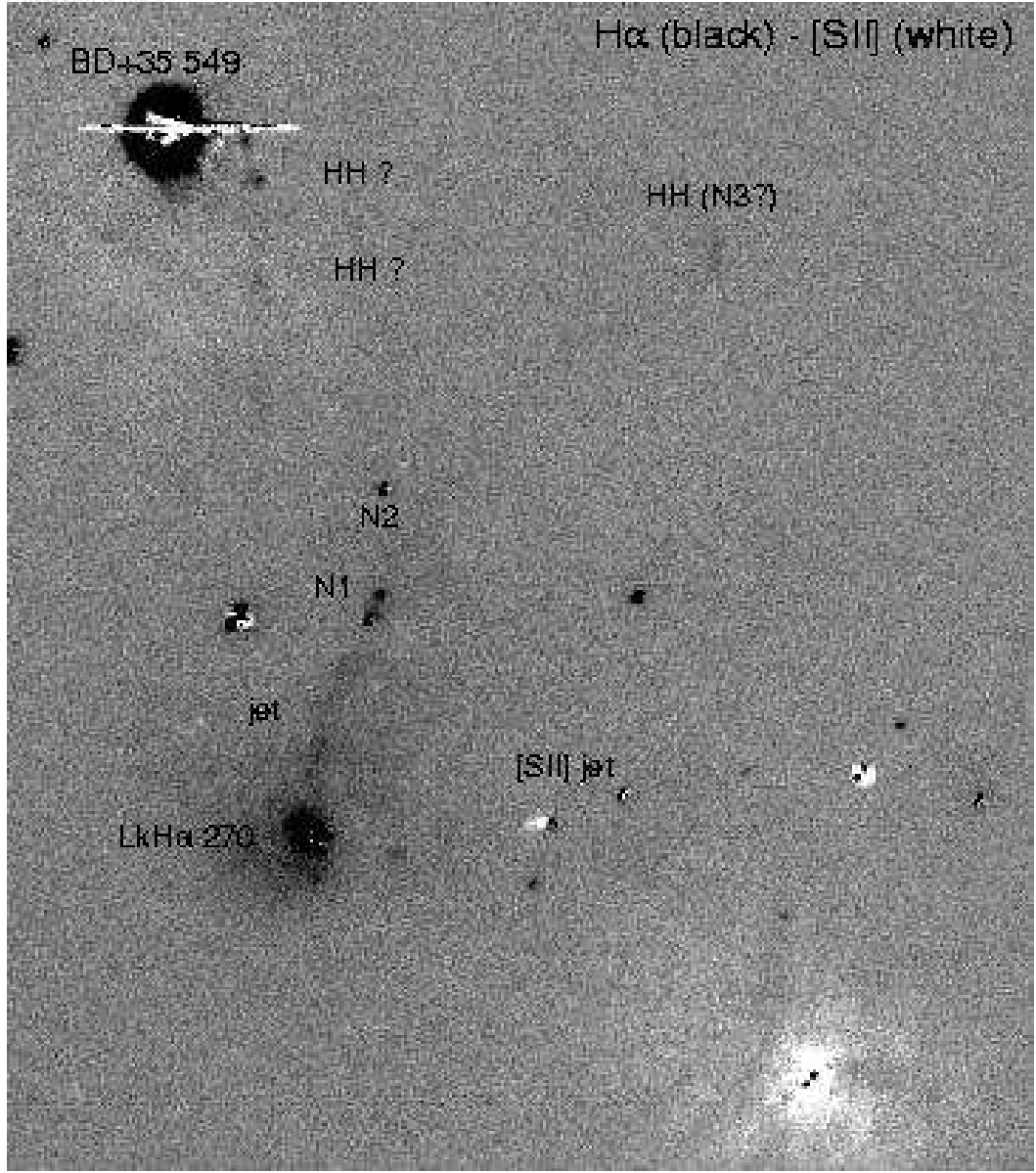}
\caption[]{
HH~335 shown in a difference image (\Ha\  - [\sii ])
with \Ha\ dominated features shown in black and
[\sii ] dominated features shown in white. 
Several additional candidate HH objects 
are also marked including a short [\sii ] jet to the
west of LkH$\alpha$270.
} 
\end{figure*}

\begin{figure*}[tbp]
\includegraphics[height=5.0in,angle=-90]{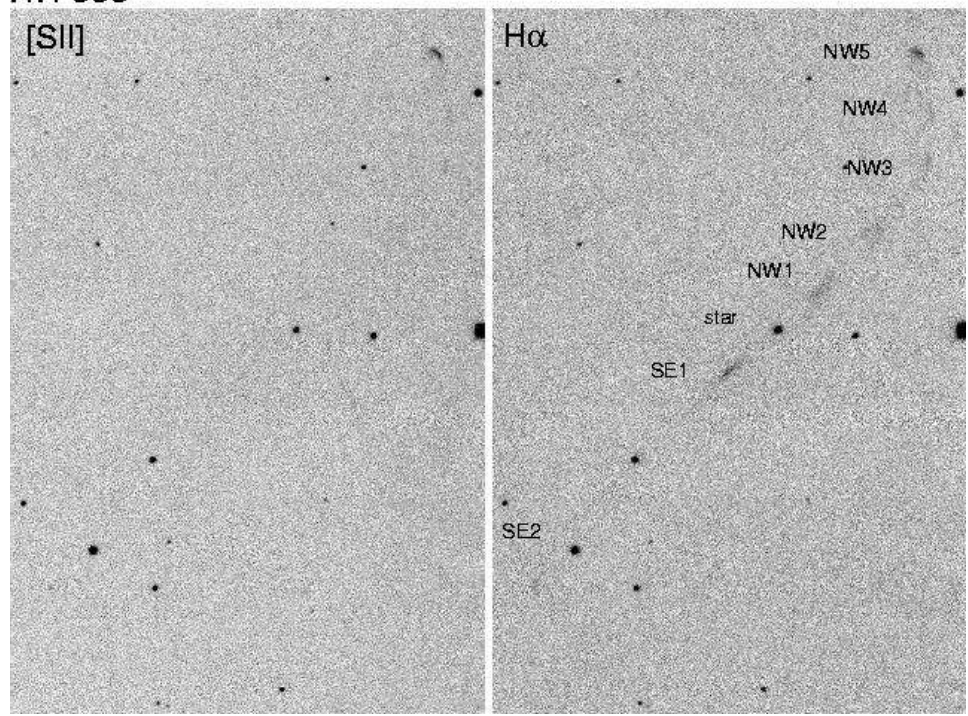}
\caption[]{
HH~336 in \Ha\ (right) and [\sii ] (left).
} 
\end{figure*}

\begin{figure*}[tbp]
\includegraphics[height=4.0in]{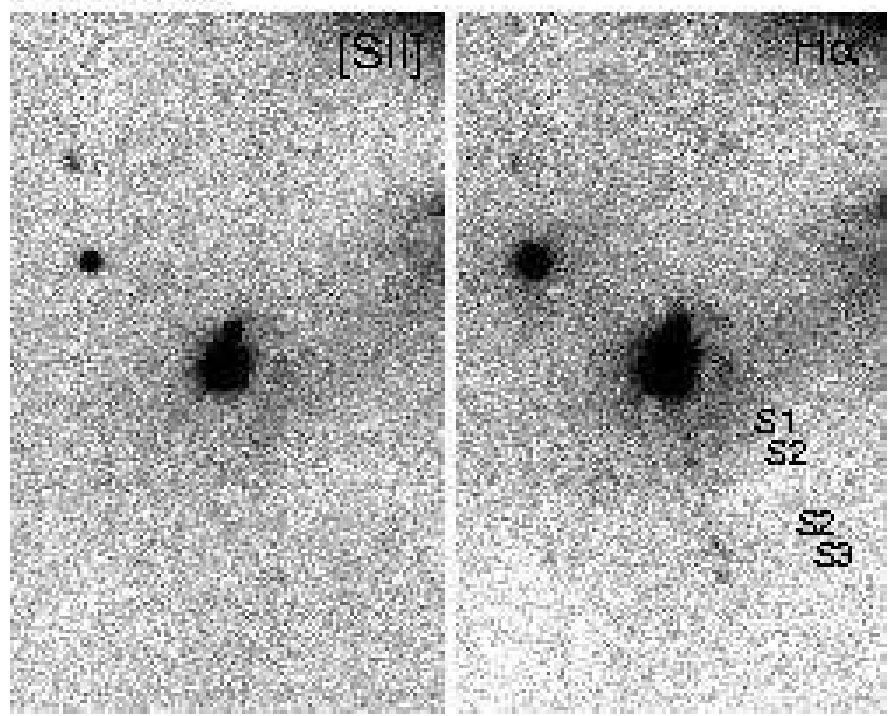}
\caption[]{
HH~495 in \Ha\ (right) and [\sii ] (left).
} 
\end{figure*}

\begin{figure*}[tbp]
\includegraphics[height=6.0in]{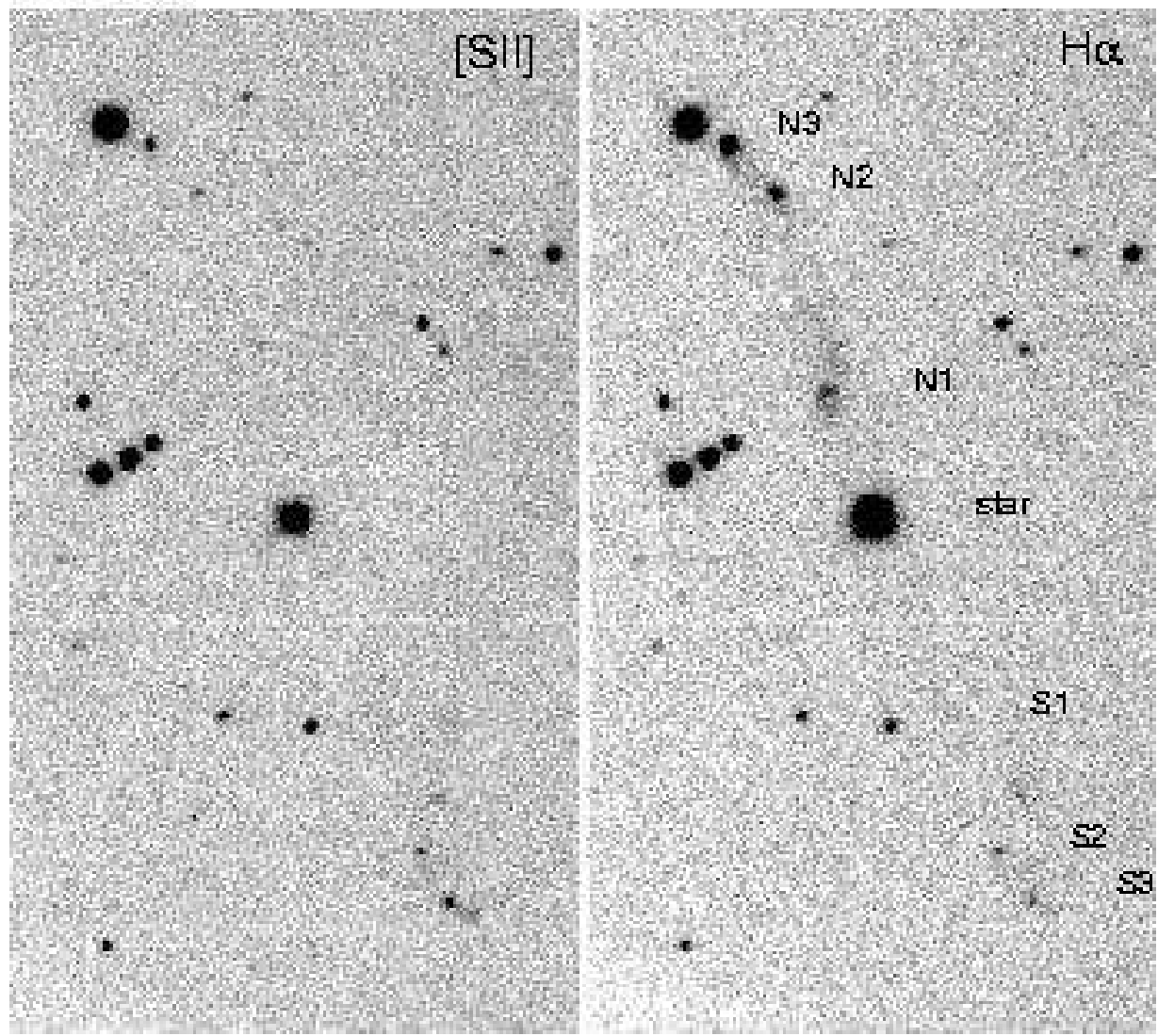}
\caption[]{
HH~497 in \Ha\ (right) and [\sii ] (left).
} 
\end{figure*}

\begin{figure*}[tbp]
\includegraphics[height=8.0in]{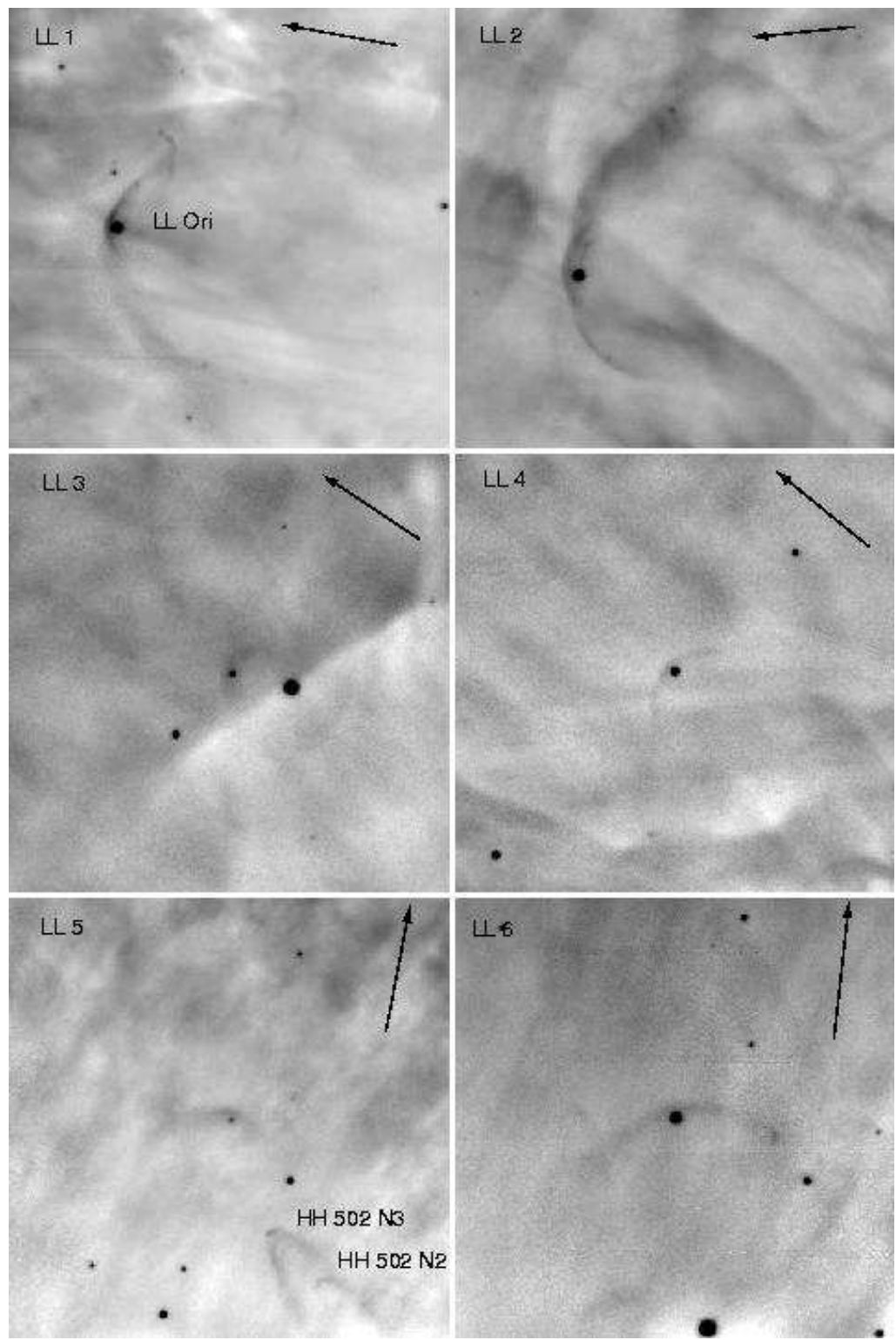}
\caption[]{
Six \Ha\ images of the LL Orionis type wind-wind collision fronts
LL~1 through LL~6 identified in the outskirts of the Orion Nebula.   
LL~1 is LL Ori.  Each image shows a 104\arcsec\ by 104\arcsec\ field 
of view with north towards the top and east towards the left.   
The arrows indicate the direction towards the Trapezium stars.  
} 
\end{figure*}

\begin{figure*}[tbp]
\includegraphics[height=3.5in]{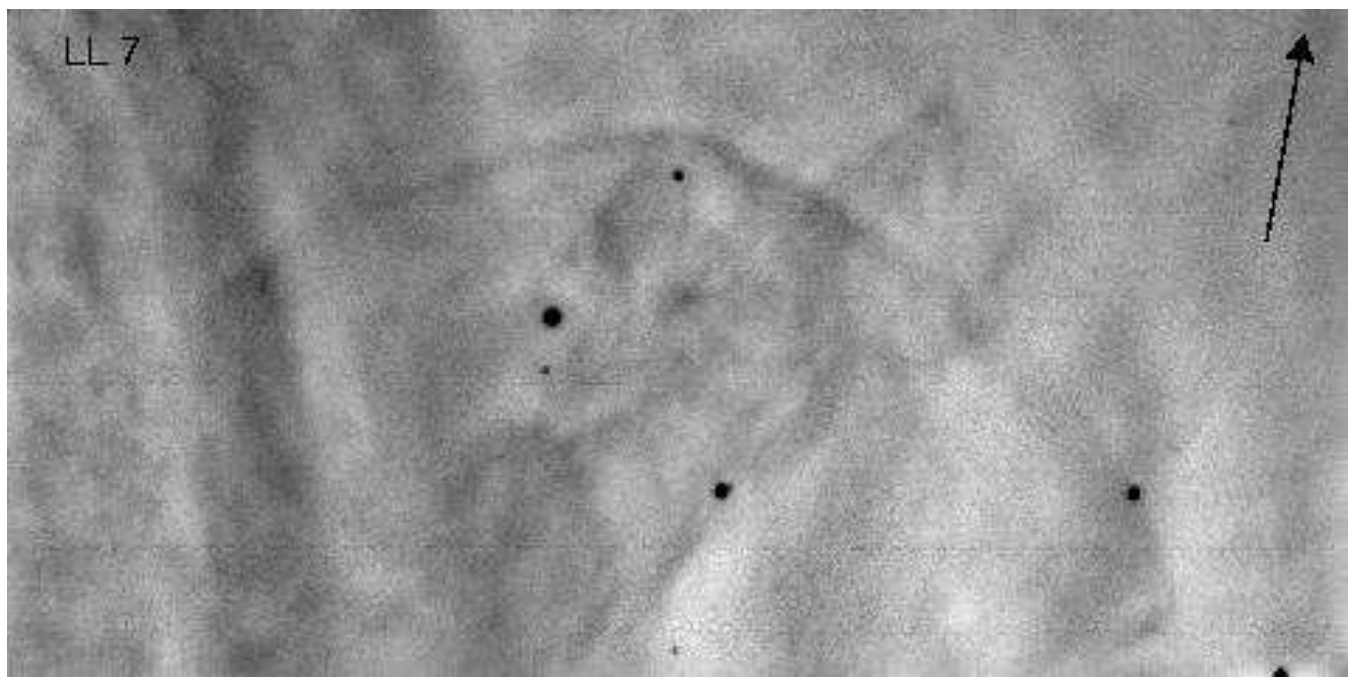}
\caption[]{
An \Ha\ image of the LL Orionis type wind-wind collision front
LL~7 south of the Trapezium. 
This image shows a 208\arcsec\ by 104\arcsec\ field of view with 
north towards the top and east towards the left.   The arrow 
indicates the direction towards the Trapezium stars.  
} 
\end{figure*}

\end{document}